\documentclass{aa}  

\usepackage{graphicx}
\usepackage{txfonts}
\usepackage{hyperref}
\hypersetup{colorlinks=true, citecolor=blue, linkcolor=blue}
\usepackage{amsmath, amsbsy, amssymb}
\usepackage{wrapfig}
\usepackage[table, svgnames, dvipsnames]{xcolor}
\usepackage{lscape}
\usepackage{float}
\usepackage{comment}
\usepackage{enumitem}
\usepackage{amsfonts}
\usepackage{subcaption}
\usepackage{fancyhdr} 
\usepackage{verbatim}
\usepackage{booktabs}
\usepackage{afterpage}
\usepackage{blindtext} 
\usepackage{multirow}
\usepackage{titlesec}
\usepackage{acronym}
\usepackage{wasysym}
\usepackage{soul}
\usepackage{array}
\usepackage{textcomp} 
\usepackage{sidecap} 
\usepackage{xfrac}
\usepackage{rotating}
\usepackage{siunitx}
\usepackage{adjustbox}
\usepackage[flushleft]{threeparttable}
\usepackage{caption} 
\usepackage{mathrsfs}

\definecolor{mypink}{RGB}{203, 175, 192}
\definecolor{operamauve}{rgb}{0.72, 0.52, 0.65}

\def\msol{\ensuremath{M_\odot}}
\def\genec{{\sc Genec}}

\begin{document} 

    \title{Rapidly rotating Population III stellar models as a source of primary nitrogen}

    \author{Sophie Tsiatsiou\inst{1}, 
            Yves Sibony\inst{1},
            Devesh Nandal\inst{1}, 
            Luca Sciarini\inst{1},    
            Yutaka Hirai\inst{2,3},
            Sylvia Ekstr\"om\inst{1}, 
            Eoin Farrell\inst{1},
            Laura Murphy\inst{9},
            Arthur Choplin\inst{4},
            Raphael Hirschi\inst{5},
            Cristina Chiappini\inst{6},
            Boyuan Liu\inst{7}, 
            Volker Bromm\inst{8}, 
            Jose Groh\inst{9}, 
            Georges Meynet\inst{1}
        }

    \institute{Department of Astronomy, University of Geneva, Chemin Pegasi 51, 1290 Versoix, Switzerland\\ \email{sofia.tsiatsiou@unige.ch}
    \and Department of Physics and Astronomy, University of Notre Dame, 225 Nieuwland Science Hall, Notre Dame, IN 46556, USA
    \and Astronomical Institute, Tohoku University, 6-3 Aoba, Aramaki, Aoba-ku, Sendai, Miyagi 980-8578, Japan
    \and Institut d'Astronomie et d'Astrophysique, Universit\'e Libre de Bruxelles, CP 226, 1050, Brussels, Belgium
    \and Astrophysics Group, Keele University, Keele, Staffordshire ST5 5BG, UK
    \and Leibniz-Institut f\"ur Astrophysik Potsdam (AIP), An der Sternwarte 16, 14482 Potsdam, Germany
    \and Institute of Astronomy, University of Cambridge, Madingley Road, Cambridge, CB3 0HA, UK
    \and Department of Astronomy, University of Texas, Austin, TX 78712, USA
    \and Independent Researcher
        }

    \date{Accepted XXX, Received YYY; in original form ZZZ}

    \abstract
    {The first stars might have been fast rotators. This would have important consequences for their radiative, mechanical, and chemical feedback.}
    {We discuss the impact of fast initial rotation on the evolution of massive Population~III models and on their nitrogen and oxygen stellar yields.}
    {We explore the evolution of Population~III stars with initial masses in the range of 9~{\msol} $\leqslant M_{\rm ini} \leqslant$ 120~{\msol}, starting with an initial rotation on the zero-age main sequence equal to 70\% of the critical one.}
    {We find that with the physics of rotation considered here, our rapidly rotating Population~III stellar models do not follow a homogeneous evolution. They lose very little mass in the case in which mechanical winds are switched on when the surface rotation becomes equal to or larger than the critical velocity. The impact on the ionising flux appears to be modest when compared to moderately rotating models. Fast rotation favours, in models with initial masses above $\sim$20~{\msol}, the appearance of a very extended intermediate convective zone around the H-burning shell during the core He-burning phase. This shell has important consequences for the sizes of the He- and CO-cores, and thus impacts the final fate of stars. Moreover, it has a strong impact on nucleosynthesis, boosting the production of primary $^{14}$N.}
    {Fast initial rotation significantly impacts the chemical feedback of Population~III stars. Observations of extremely metal-poor stars and/or starbursting regions are essential to provide constraints on the properties of the first stars.}


    \authorrunning{S. Tsiatsiou et al.}

    \maketitle

\section{Introduction} \label{sec:intro}

According to cosmological simulations, the first stars formed at redshifts of $z \sim 20 - 35$ \citep{Abel2002, Bromm2002, Yoshida2003, Ohkubo2009, Hirano2014, Susa2019} and were responsible for the first stellar radiative, chemical, and mechanical feedback in the Universe \citep[see e.g. the recent review by][and references therein]{Klessen2023}. Knowing how they evolved is therefore important for questions such as reionisation \citep{Bromm2002, Choudhury2005, Yoon2012, Sibony2022}, the first chemical enrichments \citep{LPrimN2002, Prantzos2003, Chiappini2005, CNO2006, Tominaga2007, Chiappini2008, Rollinde2009, Greif2010, Pallottini2014, Marassi2015, Sarmento2017, Hirai2018, Hirai2019, Corazza2022, Sanati2023}, the origin of carbon-enhanced metal-poor (CEMP) stars \citep{Ji2015, Clarkson2018, Liu2021CEMP, Zepeda2023}, the Population (Pop) III progenitors of long soft gamma-ray bursts \citep{Wang2012}, magnetorotational supernovae (SNe) \citep{Suwa2007, Rees2010, Suwa2011, Souza2011, MMre2012, Piro2014}, or intermediate-mass and/or merging black holes \citep{Inayoshi2016, Inayoshi2017, Regan2020, Kinu2020, Kinu2021, Liu2021, Tani2021, Santo2023, Iorio2023}, and for setting the stage for the next star formation events. In addition, recent JWST observations of high-redshift galaxies (e.g. the GLASS and CEERS surveys) have discovered a subset of N-rich galaxies \citep[see e.g.][]{Isobe2023}. In particular, it has been suggested that the intriguing z$\sim$11 source GN-z11 \citep{Cameron2023} contains a sub-population of Pop~III stars \citep{Maiolino2023}. Such Pop~III stars might be very interesting candidates for enriching these high-redshift galaxies with nitrogen over short timescales. 

The stellar feedback processes depend on characteristics of the first stellar populations, such as the initial mass function \citep{Bromm1999, Abel2002, Bromm2002}, the initial distribution of rotations, which can contain a large fraction of very rapidly rotating stars \citep{Stacy2011, Stacy2013a, HiranoBromm2018} if magnetic braking is inefficient in the absence of strong magnetic fields \citep{Hirano2018,Hirano2022,Kimura2023}, and the degree of multiplicity \citep[see e.g.][]{Stacy2010, Stacy2013b, Liu2021}, to cite only a few important features. Simulations of Pop III star formation suggest that Pop~III stars had large masses higher than 100~{\msol} due to the lack of a cooling mechanism that would prevent further fragmentation \citet{Bromm2002}. However, the exact mass range is still debated. Recent studies allow the formation of Pop III stars with masses down to 1~{\msol} and even lower \citep{Yoshida2006, Clark2011, Greif2011, Susa2013, Stacy2016, Wollenberg2020, Sugimura2020, Park2021, Prole2022, Latif2022}.

Models are also needed to find observable signatures of these primordial stars when seen as a population in high-redshift star-forming regions \citep{Schaerer2002, Schaerer2003, Raiter2010, Ma2017, Trussler2023} or when they explode in energetic, very luminous transient events \citep{Joggerst2011, Toma2011, Mesler2014, Tolstoy2016, Lee2023}. Pop~III stars are also privileged objects for studying some frontiers of physics. Since they are born in dark matter mini halos they may be impacted by the physics of the dark matter \citep{Freese2008, Taoso2008, Natarajan2009, Schleicher2009, Spolyar2009, Choplin2017, Liu2019, Liu2019dm, Liu2020}. Also, some cosmological theories predict that fundamental constants, such as the fine structure constant, may evolve as a function of time over cosmic history. Pop~III stellar models have been used to test the impact of different values of the fine structure constant on the stellar nucleosynthesis in the very early Universe \citep{Ekstrom2010, Huang2019, Mori2020}. 

In the last two decades or so, evolutionary features of non-rotating Pop~III stellar models have been studied by, among others, \citet{Marigo2001, Chieffi2001, Chieffi2002a, Chieffi2002b, Ohkubo2009, Bahena2010, HegerWoosley2010}. Work has also been done to study the impact of close binary evolution of Pop~III stars \citep[see e.g.][]{Chen2015, Tsai2023}. The impact of convective interactions between H- and He-burning regions in massive Pop~III stellar models has been studied by \citet{Clarkson2021}.

\citet{Sylvia2008} explored the impact of rotation on the evolution and chemical feedback of Pop~III stars. \citet{LauraGrids} performed a similar study as \citet{Sylvia2008}, but adopted a different setting for the diffusion coefficients describing the mixing of the elements by rotation. More recently, \citet{Aryan2023} used MESA to compute the evolution of Pop~III rotating 25~{\msol} models.

Rotation deeply impacts nucleosynthesis. For instance, \citet{LPrimN2002, Chiappini2006, Sylvia2008} have shown that the diffusion of He-burning products into the H-burning shell induced by shear turbulence boosts the production of isotopes such as $^{13}$C and $^{14}$N. Further works have shown that the diffusion of nitrogen into the He-burning core boosts the production of $^{19}$F, $^{22}$Ne, s-, and p-process elements \citep{CNO2002, CNO2006, Chiappini2008, Pigna2008, Cesc2013, Frisch2016, Limongi2018, Chopl2018, Chopl2022} and given them the status (at least in part) of primary elements. Let us recall that a primary element is an element whose production does not scale with the initial metallicity but is produced from the transformation of hydrogen and helium \citep[see e.g. the book by][]{Matt2001}. An interesting example is the production of primary nitrogen in metal-poor rotating stars \citep{CNO2002}. The turbulence induced by rotation triggers mixing between the H- and He-burning zones, allowing some carbon and oxygen produced in the He-burning zone to diffuse into the H-burning zone. In the H-burning shell, carbon and oxygen are transformed into nitrogen by the action of the carbon–nitrogen–oxygen (CNO) cycle. This nitrogen qualifies as primary because it is produced from carbon and oxygen produced by the star and not from carbon and oxygen that were present in the star at its formation. 

In past works, it has been shown that for a given initial mass, metallicity, and rotation, the primary nitrogen production only occurs at low metallicity, typically at metallicities between $Z \sim 10^{-8}$ and $10^{-4}$, where $Z$ is the mass fraction of the abundances of all the elements heavier than helium \citep[see e.g.][]{CNO2006}. The reason why this mixing is efficient only at low metallicities is mainly due to the fact that the lack of CNO elements at the beginning of the evolution makes the H-burning occur at higher temperatures than in more metal-rich stars. As a consequence, the H- and He-burning zones are less distant from each other in metal-poor stars, and the entropy gradients are weaker, making any mixing either by convection or by diffusion easier.

Previous modelling of zero-metallicity stars has suggested that Pop~III stars will produce significantly less primary nitrogen than a model with a very small initial metal content (assuming that models start with a velocity at the zero-age main sequence (ZAMS), $\upsilon_{\rm ini}$, corresponding to the same constant fraction of the critical velocity\footnote{The critical velocity is the velocity at the equator where the gravity is balanced by the centrifugal force.}, $\upsilon_{\rm crit}$, at all metallicities). This occurs because some degree of differential rotation is required to trigger the mixing. The differential rotation is mainly produced by the contraction of the core at the end of the main sequence (MS) phase. Contraction in a Pop~III star is not as great as in an even slightly metal-enriched star, because as we explained before, the temperature of He-burning is not much greater than the temperature in the H-burning core. A modest core contraction is therefore sufficient to raise the core temperatures to the level needed to achieve radiative equilibrium. Furthermore, the opacity in the envelope is lower, facilitating the evacuation of the energy released by contraction from the envelope. Thus, less energy is used to expand the envelope, which also suppresses differential rotation. The star remains in the blue part of the Hertzsprung-Russel diagram (HRD) for the whole core He-burning phase, or at least a substantial part of it\footnote{\citet{Farrell2022} has performed very interesting numerical experiments showing how artificially preventing the contraction actually prevents the crossing of the Hertzsprung-Russel (HR) gap. Actually, things are more complex in the sense that some contraction always occurs and other factors such as the distribution of helium around the H-burning shell are important too.}.

In this paper, we explore Pop~III stellar models rotating faster on the ZAMS than in our previous works and how significant the production of primary nitrogen can be in those. Therefore, instead of considering an initial velocity of $\upsilon_{\rm ini}=0.4\,\upsilon_{\rm crit}$, as in \citet{LauraGrids}, we consider $\upsilon_{\rm ini}=0.7\,\upsilon_{\rm crit}$. In Sect.~\ref{sec:2}, we describe the ingredients of the stellar models. Some effects of fast rotation on the evolution of primordial stars are discussed in Sect.~\ref{sec:3}. A brief presentation of the small grid computed in this work is given in Sect.~\ref{sec:4}. We discuss the impact of fast rotation on the primary nitrogen production in Sect.~\ref{sec:5}. Sect.~\ref{sec:6} compares the yields obtained from models at different metallicities by different authors. The impact of the present yields on a simple chemical evolution model is also presented. The conclusions are given in Sect.~\ref{sec:7}.

\section{Ingredients of stellar models} \label{sec:2}

The rapidly rotating Pop~III models have been computed with the Geneva stellar evolution code \citep[{\genec}, see a detailed description in][]{Eggenberger2008} with the same physical ingredients as the Pop~III models published by \citet{LauraGrids} and models at higher metallicities by \citet{Sylvia2012, Cyril2013, Groh2019, Eggenberger2021, Yusof2022}. The reader is invited to refer to those papers for the details of the input physics. Only a few relevant ingredients are recalled here.

The initial composition considered is $X=0.7516$, $Y=0.2484$, and $Z=0$, with $X$, $Y$, and $Z$ the mass fractions of hydrogen, helium, and heavy elements, respectively. The primordial $^{4}$He abundance comes from predicted values from the Big Bang nucleosynthesis \citep{He-abund}.

We explore the effects of fast rotation on the evolution of massive stars with similar masses to those considered in \citet{LauraGrids}. We restricted masses here to between 9 and 120~{\msol}. More massive star models have been recently explored using {\genec} in \citet{Martinet2023} and \citet{THREE2023, CRIT2023}, although not with fast rotation.

Rotation was implemented according to the stellar rotation theory developed by \citet{Zahn1992}. The expressions describing the vertical and horizontal shear diffusion coefficients are from \citet{Maeder1997} and \citet{ChaboyerZahn1992}, respectively. Magnetic fields are not accounted for in the present models.

We did not account for any mass loss due to line-driven winds, as this was discussed in \citet{LauraGrids}. Even though there are many other processes that may allow Pop~III stars to lose mass \citep[see e.g. the discussion in the introduction of][]{Liu2021CEMP} -- for example, pulsation-driven mass loss \citep{Volpato2023}, mass loss during a mass transfer in a close binary, and mechanical mass loss -- in this work only mechanical mass loss has been taken into account. Mechanical mass loss occurs when the surface rotation reaches a critical velocity. We estimated the mass lost mechanically as the mass that needs to be removed from the star in order to keep its surface rotation sub-critical. A detailed account of the exact numerical procedure is given in \citet[Sect. 2.2]{Georgy_Be}.

The critical velocity is defined as the surface rotation such that the centrifugal acceleration at the equator balances the gravity there. In this study, we followed the expression given by \citet{OGlimit2000}, which accounts for the fact that the critical velocity varies when the luminosity of the star is near the Eddington luminosity (the so-called $\Omega\Gamma$ limit, where $\Omega$ is the angular velocity and $\Gamma$ the Eddington factor). The critical rotation has two roots depending on the Eddington factor:
\begin{equation}
    \upsilon_{\rm crit} =
    \begin{cases}
      \sqrt{\frac{2}{3}\frac{GM}{R_\text{p,crit}}} & , \text{$\Gamma < 0.693$}\\
      \sqrt{\frac{3}{2}\upsilon_{\rm crit,1}\frac{R_\text{e,crit}}{R_\text{p,crit}}\frac{1-\Gamma}{{V}'(\omega))}} & , \text{$\Gamma > 0.693$}\\
    \end{cases} 
,\end{equation}
where $R_\text{p,crit} = \frac{2}{3}R_\text{e,crit}$, $R_\text{p,crit}$ and $R_\text{e,crit}$ are the polar and equatorial radius at critical velocity, respectively, $V'(\omega)$ is the ratio between the actual volume of the deformed rotating star to a spherical volume with a radius equal to $R_\text{p,crit}$, and $\omega$ is the ratio of the surface angular velocity to the critical angular velocity (given by the first root in the equation just above).

The limits of the convective cores were determined using the Schwarzschild criterion, with a step overshooting parameter of $d_{\rm over}/H_P = 0.1$ for the core H- and He-burning phase ($d_{\rm over}$ is the spatial extension of the convective core above the radius given by the Schwarzschild's limit and $H_P$ is the pressure scale height at Schwarzschild's boundary).

We complement the grid by \citet{LauraGrids} for specific masses (9, 20, 60, 85, and 120~{\msol}) with models that have an initial rotation rate of $\upsilon_{\rm ini}/\upsilon_{\rm crit}=0.7$. This gives initial rotations on the ZAMS that are between 650 and 1200 km/s (see column 2 in Table~\ref{table:models_details-2}\footnote{Keeping constant values for $\upsilon_{\rm ini}/\upsilon_{\rm crit}$ for all the initial masses implies a faster initial surface velocity for more massive stars.}). At first sight, these rotations may appear to have quite extreme values, at least well outside the range of surface velocities measured in the present-day Universe. However, the perception is different if we assume that stars begin their evolution on the ZAMS with a total angular momentum content that is kept constant for all metallicities. Since Pop~III stars of a given mass have a much smaller momentum of inertia than solar metallicity stars, they naturally have faster rotations. To give a numerical example, the total angular momentum of a solar metallicity 9~{\msol} model with an initial surface velocity equal to 40\% of the critical one ($264\,{\rm km}/{\rm s}$) is $0.072\times10^{53}\,{\rm g}\,{\rm cm}^{2}/{\rm s}$ \citep[models by][]{Sylvia2012}. The total angular momentum of a Pop~III 9~{\msol} model with an initial surface velocity equal to 70\% of the critical one ($649\,{\rm km}/{\rm s}$) is $0.085\times10^{53}\,{\rm g}\,{\rm cm}^{2}/{\rm s}$. The angular momenta are not very different, but the rotation rates are. 

\section{Impact of fast rotation on primordial stars} \label{sec:3}

In this section we explore the effect of fast rotation on primordial stars, focusing on transport processes, surface He-enrichments, rotations at the critical limit, and rotational mixing due to an effect that we call CNO shell boost. 

\subsection{Transport processes} \label{sec:31}

    \begin{figure*}
    \centering
    \includegraphics[width=1\textwidth]{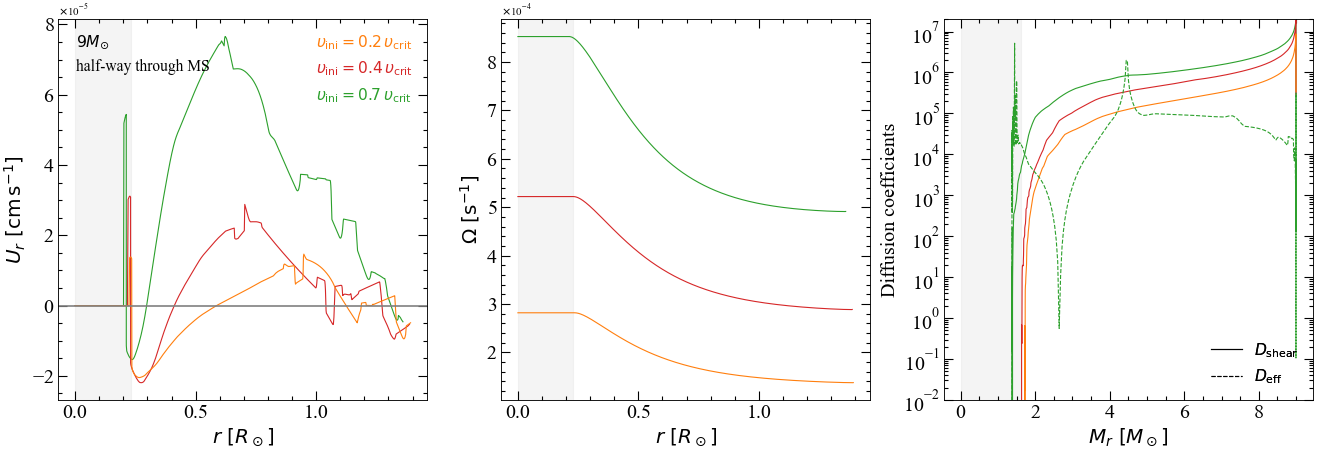}
    \caption{Internal profiles as a function of the radius and the Lagrangian mass coordinate inside 9~{\msol} models halfway through the MS ($X_{c}=0.35$) for different initial rotations: $\upsilon_{\rm ini}/\upsilon_{\rm crit}=$ 0.2 (orange), 0.4 (red) \citep{LauraGrids}, and 0.7 (green). The grey-shaded zone is the convective core of the moderately rotating model.
    {\it Left panel:} Radial component of the meridional circulation velocity, $U_r$. 
    {\it Middle panel:} Angular velocity, $\Omega$. 
    {\it Right panel:} The solid curve shows the diffusion coefficient due to shear turbulence, $D_{\rm shear}$. The dotted curve shows the effective diffusion coefficient, $D_{\rm eff}$ (y-axis in logarithmic scale and in units of ${\rm cm}^2/{\rm s}$). For clarity, only the curve for $D_{\rm eff}$ corresponding to $\upsilon_{\rm ini}=0.7\,\upsilon_{\rm crit}$ is shown. In the rest of the radiative zone, $D_{\rm eff}$ is in general much smaller than $D_{\rm shear}$.} 
    \label{fig:9-Xc035}
    \end{figure*}

Figure~\ref{fig:9-Xc035} shows the part of the radial component of the meridional circulation velocity\footnote{The radial component of the meridional circulation velocity is given by $u_r(r, \theta)=U_r(r)P_2(\cos \theta)$, where $u_r(r, \theta)$ is the radial component of the physical velocity, $\theta$ is the co-latitude, and $P_2$ the second Legendre polynomial.}, $U_{r}$, that depends on the radial coordinate, $r$, and the angular velocity in 9~{\msol} stellar models halfway through the MS phase (e.g. when the mass fraction of hydrogen at the centre is $X_{\rm c}=0.35$) for three different initial rotations. Increasing the initial rotation significantly enhances $U_r$ both in the radiative envelope (at radial coordinates equal to 0.6-0.7 $R_\odot$) and just above the convective core (at around 0.2 $R_\odot$). In the models sketched in Fig.~\ref{fig:9-Xc035}, we can identify at least three changes of sign for $U_{r}$ (when $U_{r}$ is positive, the meridional currents transport the angular momentum from the outer regions to the inner ones, and the reverse is true when it is negative). We see that just above the core, and further out in the radiative envelope, there are regions where the meridional currents transport the angular momentum inwards, reinforcing there the gradient of the angular velocity, $\Omega$. This behaviour cannot be modelled by approximating the angular momentum transport by meridional currents as a diffusive process. Indeed, a diffusive process will always tend to smooth any $\Omega-$gradient.

The interactions between the meridional currents, the effects induced by the changes of structure, and the diffusion by shear on the angular momentum distribution result in the variation of the angular velocity with the radius plotted in the middle panel of Fig.~\ref{fig:9-Xc035}. It shows that the gradients of $\Omega$ are steeper in the rapidly rotating models, which implies more chemical mixing by shear. This is consistent with the behaviour of these models in the HRD (see Fig.~\ref{fig:HRDrot}), where indeed the rapidly rotating models follow, once the chemical mixing has had time to be significant, a bluer and more luminous track that is indicative of a more efficient chemical mixing. The steepness of the $\Omega-$gradient is caused by the core contraction and the transport of the angular momentum. As has been discussed in many previous papers \citep[see e.g.][]{MM2000}, in models without any magnetic fields, angular momentum is mainly transported by meridional currents. In regions where $U_{r}$ is positive, the angular momentum is transported from the outer layers into the inner ones, reinforcing the gradient produced by the core contraction and envelope expansion. As can be seen from the left panel of Fig.~\ref{fig:9-Xc035}, $U_{r}$ is positive in a large part of the radiative envelope and reaches larger values in the faster rotating model, thus making the $\Omega-$gradient steeper. Such a process cannot be reproduced when the transport of the angular momentum by the meridional currents is described by a diffusive equation, as is done in some codes.

In the 9~{\msol} model, the mixing of the elements is mainly governed by the vertical shear diffusion coefficient ($D_{\rm shear}$), except in a small region above the convective core where the diffusion coefficients describing the results of the action of both the meridional currents and of the horizontal shear turbulence ($D_{\rm eff}$) dominate (see the right panel of Fig.~\ref{fig:9-Xc035}). 
In contrast, in more massive models (typically for the 60~{\msol} model and above), the main driver for the transport of the chemical species is $D_{\rm eff}$ (see the lower right panel of Fig.~\ref{fig:120-Xc035}). The diffusion coefficient, $D_{\rm eff}$, scales with the meridional velocity \citep{ChaboyerZahn1992}, which in turn scales as the inverse of the density. The latter scaling implies that for a given rotation the meridional velocities are in general larger in less dense, more massive stars. As a numerical example, the maximum value is nearly $8\times 10^{-5}\,{\rm cm}/{\rm s}$ in the rapidly rotating 9~{\msol} model plotted in Fig.~\ref{fig:9-Xc035}. In the corresponding 120~{\msol} model, the absolute value is of the order of $1\,{\rm cm}/{\rm s}$, and hence four orders of magnitude larger (see the upper left panel of Fig.~\ref{fig:120-Xc035}). Moreover, it is a negative value, indicating that it transports angular momentum from the inner layers to the outer regions. This transport appears to be more efficient in rapidly rotating stars.

\subsection{Surface He-enrichment} \label{sec:32}

    \begin{figure}
    \centering
    \includegraphics[width=0.45\textwidth]{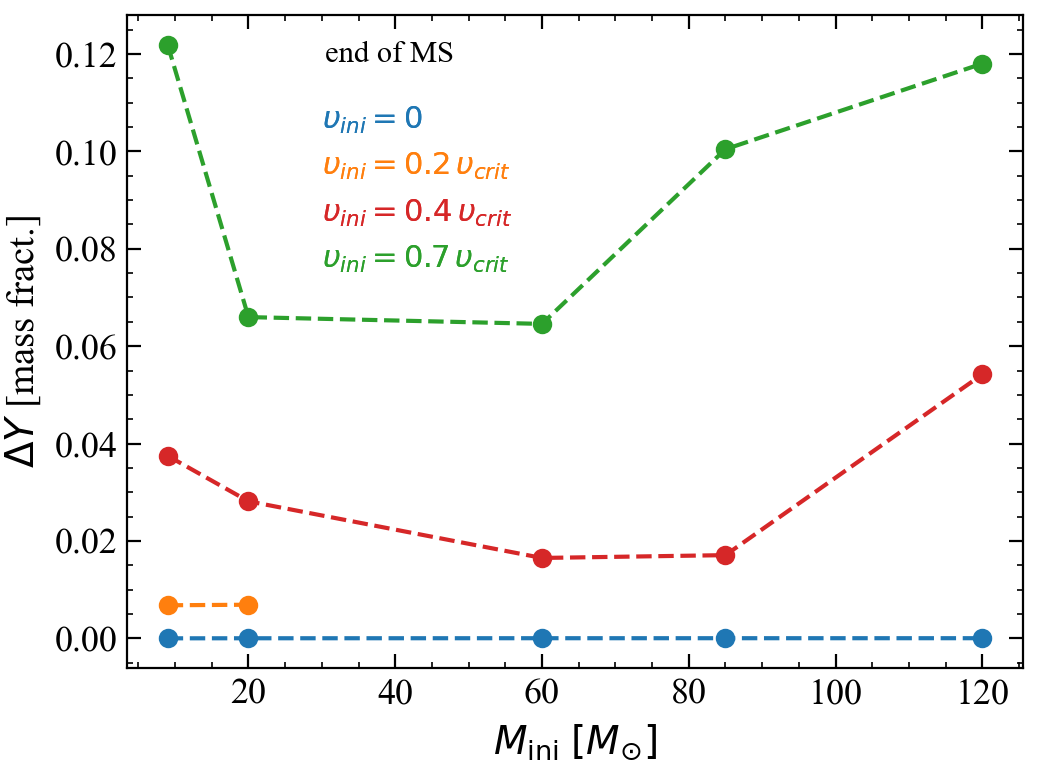}
    \caption{Surface $^{4}$He-enrichment at the end of the MS phase (see text) as a function of the initial mass for all the Pop~III models presented in Table~\ref{table:models_details-2}.}
    \label{fig:DY}
    \end{figure} 

In Fig.~\ref{fig:DY} we show the surface $^{4}$He-enrichment at the end of the MS phase for different initial masses and rotations. The quantity plotted is $\Delta Y = Y_{\rm surf}-Y_{\rm ini}$, where $Y_{\rm ini}=0.2484$ is the initial abundance of $^{4}$He. The most rapidly rotating model (see the green curve) achieves larger surface He-enrichment than the less rapidly rotating ones at the end of the MS phase. Interestingly, the curves for a fixed value of $\upsilon_{\rm ini}/\upsilon_{\rm crit}$ are non-monotonic. They present a concave shape (see the curves for the initial rotation, $0.4\,\upsilon_{\rm crit}$ and $0.7\,\upsilon_{\rm crit}$). This shape is due to the following effects: on the one hand, the radius of a star with a lower initial mass is smaller than the one of a more massive star, implying that, for a given shear diffusion coefficient, the mixing timescale is smaller in the case of a lower initial mass star (the timescale of mixing due to a diffusion coefficient, $D$, in a zone of spatial extent, $R$, scales as $R^2/D$). On the other hand, the total diffusion coefficient that is the sum of $D_{\rm shear}$ and $D_{\rm eff}$ is in general larger in more massive stars (compare the right panel of Fig.~\ref{fig:9-Xc035} with the two lower panels of Fig.~\ref{fig:120-Xc035}). This makes the mixing timescale shorter over a given distance. These effects, together with the different sizes of the convective cores in the different initial mass models, contribute to producing the curves shown in Fig.~\ref{fig:DY}.

\subsection{Population III stars at the critical limit} \label{sec:33}

    \begin{figure}
    \centering
    \includegraphics[width=0.48\textwidth]{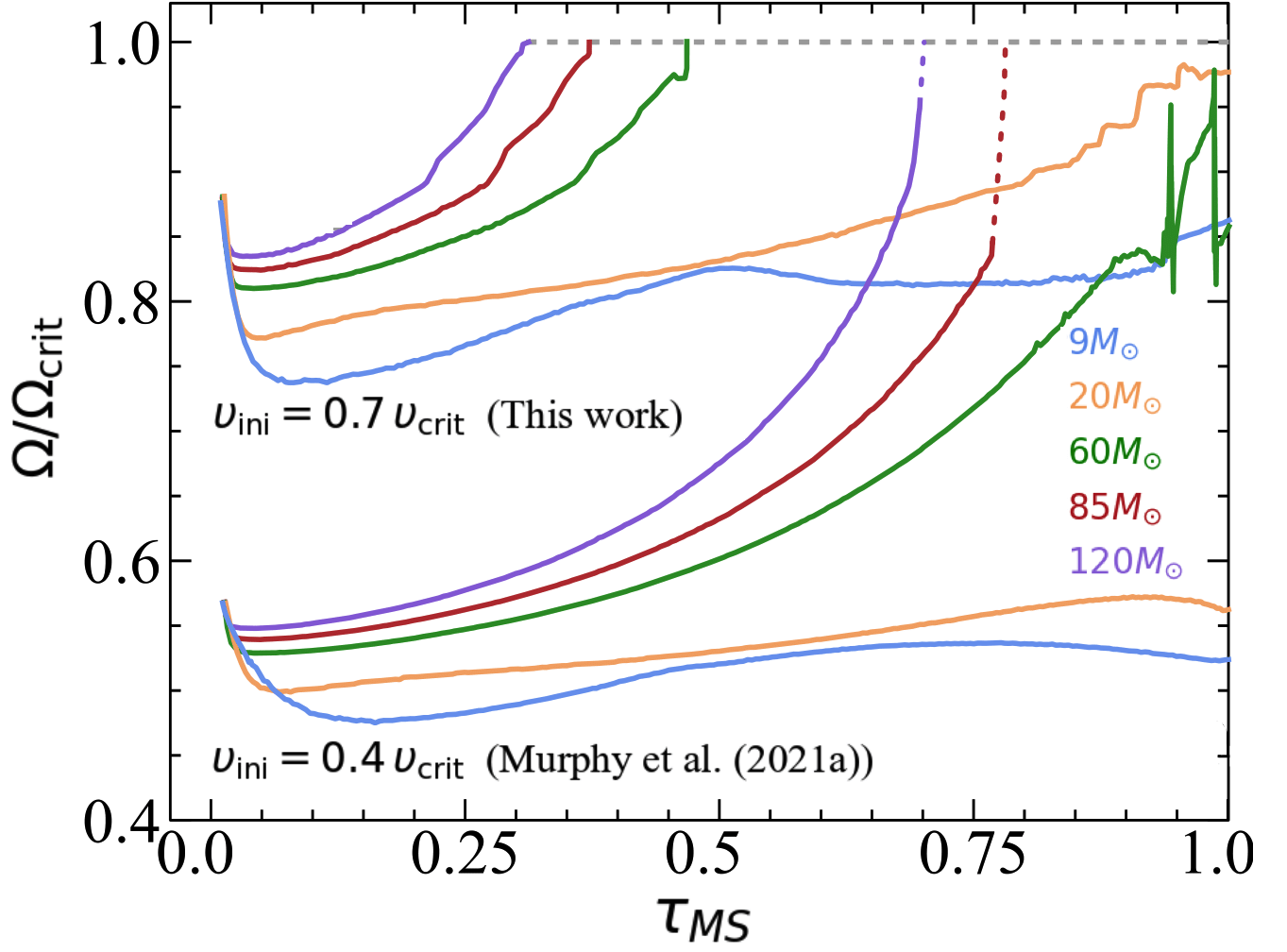}
    \caption{Evolution of the ratio between the surface angular velocity and the critical one as a function of the fraction of the total MS lifetime ($\tau_{\rm MS}$), for models within the mass range of 9~{\msol} $\leqslant M_{\rm ini} \leqslant$ 120~{\msol} with different initial rotations. The dashed grey line indicates the critical limit. The moderately rotating models are from \citet{LauraGrids}.}
    \label{fig:massloss}
    \end{figure}

Figure~\ref{fig:massloss} shows the evolution of the ratio between the surface angular velocity and the critical one during the MS lifetime for models with moderate and fast rotation. The rapidly rotating models reach the critical velocity at an earlier stage of the evolution, during the core H-burning phase (the models with $M_{\rm ini} \geqslant 60\,{\msol}$ spend more than 50\% of their MS lifetime at the critical limit). However, the maximum mass that these models can lose via this process is only at most 2\% of their initial mass, so the present models can be considered to be evolving at a nearly constant mass. We note that it appears plausible that such stars would be surrounded by a decretion disc as Oe-Be stars observed in the present-day Universe. They would be He-rich stars according to the present models.

\subsection{CNO shell boost} \label{sec:34}

    \begin{figure}
    \centering
    \includegraphics[width=0.45\textwidth]{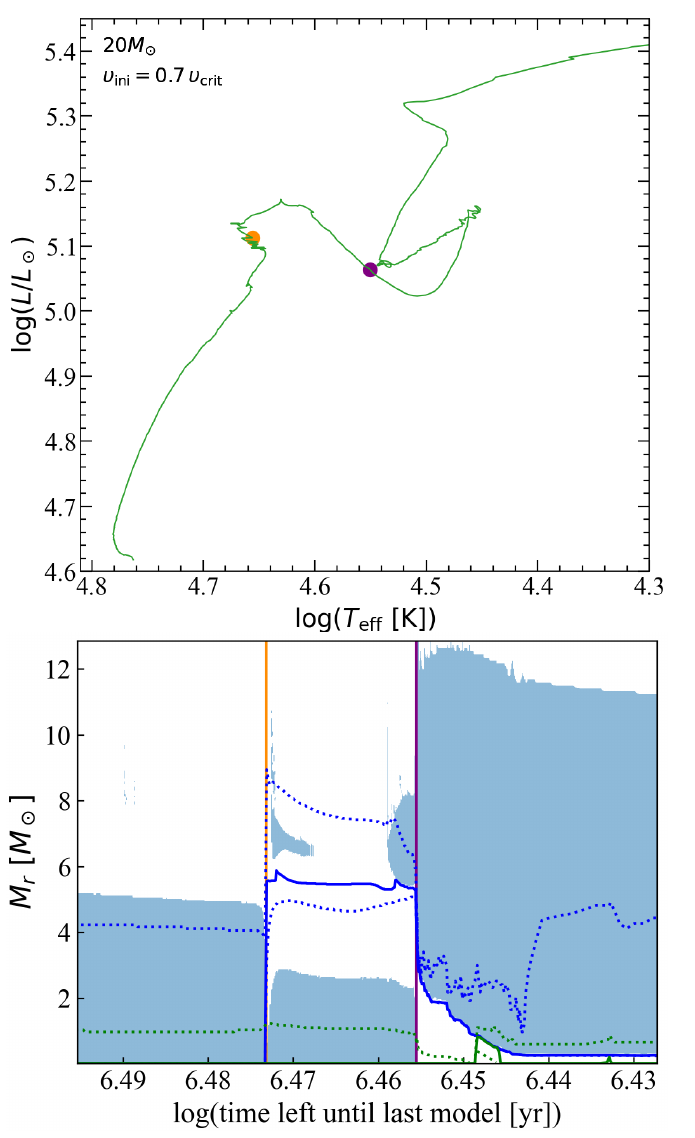}
    \caption{Evolutionary track for the 20~{\msol} rapidly rotating model ({\it upper panel}) and Kippenhahn diagram ({\it lower panel}). The orange and purple points overplotted on the evolutionary track in the upper panel have an internal structure shown in the lower panel at the two times indicated by the vertical orange and purple lines, respectively. The blue-shaded zones are the convective areas, while the white zones are the radiative ones. The blue and green curves show the mass coordinates where hydrogen and helium are burning, respectively. Furthermore, the solid curves correspond to the peak of the energy generation rate and the dashed ones to 10\% of the peak energy generation rate for each burning phase.}
    \label{fig:kippen-points20}
    \end{figure}

\citet{Sylvia2008} showed that rotational mixing can drive an H-shell boost due to the injection (by rotational mixing and/or convection) of significant amounts of carbon and oxygen from the He-burning core. This boosts the energy generation in the shell and may sometimes give birth to a very extended intermediate convective zone attached to the H-burning shell. The boost derives from the fact that the nuclear reactions of $pp$-chains are taken over by the CNO cycle, with a very different dependence on temperature. In some extreme cases, the H-shell may even become for a time the main source of energy in the star, quenching the energy produced in the He-burning core. When the mixing is gentle and progressive, the star has time to readjust and evolve in a continuous, smooth way. When the mixing is much more rapid, it may produce a strong and abrupt change of the structure like the extreme case sketched in Fig.~\ref{fig:kippen-points20}. In the following, we designate such an event a CNO shell boost.

The upper panel of Fig.~\ref{fig:kippen-points20} shows the evolutionary track in the HRD of the rapidly rotating 20~{\msol} model. The orange point along the track indicates the end of the core H-burning phase and the beginning of the core He-burning phase. The structure of the star at that stage is given by the vertical orange line in the Kippenhahn diagram shown below. The core He-burning occurs in a convective core with a mass of about 3~{\msol}, but suddenly the convective core disappears when, due to the process described above, carbon and oxygen are injected into the H-burning shell. This increases the nuclear energy produced in the shell. An intermediate convective shell appears that facilitates still more the feeding of the H-shell in carbon and oxygen. A CNO shell boost occurs. It produces such a strong release of energy that the intermediate convective zone extends from the mass coordinate 2 up to 12~{\msol}. The purple point along the evolutionary track shows the position in the HRD where the CNO shell boost occurs, and the corresponding internal structure is indicated by the vertical purple line in the Kippenhahn diagram. From that stage onwards, the He-burning convective core disappears for a while. The star is mainly powered by H-burning (in a very extended shell). This produces a `new' MS track in the HRD.

A similar figure for the 60~{\msol} is shown in Fig.~\ref{fig:kippen-points60}. Qualitatively, we have a similar evolution as the 20~{\msol} model. After a phase of core He-burning (between time coordinates around 5.52 and 5.32 in a logarithm of the remaining time in years) during which no convective shell is associated with the H-burning shell, an intermediate convective shell appears above the H-burning region. This never succeeds in quenching the nuclear reaction in the core but significantly reduces the size of the He-burning convective core. Such a feature in very massive stars may prevent the star from entering the pair-instability regime and may have importance in determining the lower limit of the black hole mass gap \citep{Eoin2021}. 

\section{The rapidly rotating grid} \label{sec:4}

    \begin{figure*}
    \centering
    \includegraphics[width=0.65\textwidth]{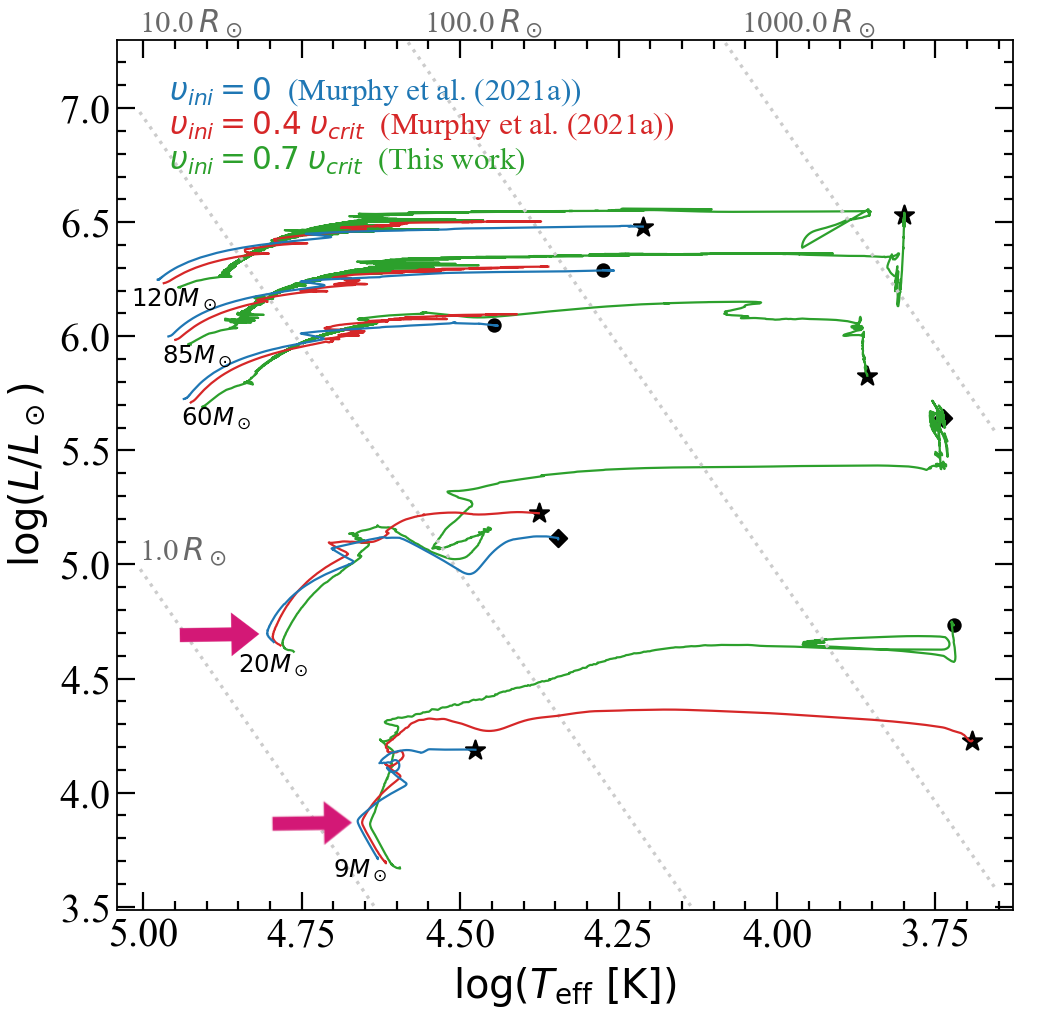}
    \caption{HRD of the Pop~III models from \citet{LauraGrids} (blue: $\upsilon_{\rm ini}=0$, red: $\upsilon_{\rm ini}=0.4\,\upsilon_{\rm crit}$) and the rapidly rotating models computed for this work (green: $\upsilon_{\rm ini}=0.7\,\upsilon_{\rm crit}$). The symbols drawn at the end of the tracks mark the final evolutionary stage attained: a star for models that have completed the core He-burning phase, a circle for models that are either in the core C-burning phase or at the end of it, and a diamond for models that are after the core C-burning phase. The models without any symbol were stopped in the core He-burning phase. Iso-radius lines are drawn in dotted grey. The pink arrows indicate the transition between the {\it pp-contraction} phase and the {\it pp-CNO-sustained} phase.}
    \label{fig:HRDrot}
    \end{figure*}

    \begin{table}
    \scriptsize
    \centering
    \caption{Properties of Pop~III models at the end of the core H- and He-burning phases with different initial surface velocities. Models with $\upsilon_{\rm ini}/\upsilon_{\rm crit}=$ 0, 0.2, and 0.4 are from \citet{LauraGrids}. Models that have not finished the core He-burning phase are indicated with a horizontal dash.}
    \addtolength{\tabcolsep}{-2pt}
    \begin{tabular}{cc|cccc|cccc}
    \hline																	
    \hline \noalign{\smallskip}	
	   &		&	\multicolumn{4}{c|}{end MS}	 &	\multicolumn{4}{c}{end He-burning}	\\
    \hline \noalign{\smallskip}		
    $M_{\rm ini}$ & $\upsilon_{\rm surf}$	&	$\tau_{\rm H}$	& $\upsilon/\upsilon_{\rm crit}$	&	$\upsilon_{\rm surf}$	&	 {Y$_{\rm surf}$}	&	$\tau_{\rm He}$	&	$\upsilon/\upsilon_{\rm crit}$	&	$\upsilon_{\rm surf}$	&	 {Y$_{\rm surf}$}	\\
    $[${\msol}]	&	[km/s]	&	[Myrs]	&	&	[km/s]	&	&	[Myrs]	&	&	[km/s]	& \\
    \hline																			
    \hline \noalign{\smallskip}		
    \rowcolor{mypink}
    \multicolumn{10}{c}{$\upsilon_{\rm ini}=0$}	\\
    \noalign{\smallskip}
    9	&	0	&	17.7&	0	&	0	&	0.248	&	1.9	&	0	&	0	&	0.248	\\ [1ex]
    20	&	0	&	9.5	&	0	&	0	&	0.248	&	0.6	&	0	&	0	&	0.248	\\ [1ex]
    60	&	0	&	3.7	&	0	&	0	&	0.248	&	0.3	&	0	&	0	&	0.248	\\ [1ex]
    85	&	0	&	3.1	&	0	&	0	&	0.248	&	0.3	&	0	&	0	&	0.248	\\ [1ex]
    120	&	0	&	2.7	&	0	&	0	&	0.248	&	0.3	&	0	&	0	&	0.248	\\ [1ex]
    \rowcolor{mypink}
    \multicolumn{10}{c}{$\upsilon_{\rm ini}=0.2\,\upsilon_{\rm crit}$}	\\
    \noalign{\smallskip}	
    9	&	23.1	&	19.9	&	0.18	&	124	&	0.255	&	3.3	&	0.10	&	41	&	0.260	\\ [1ex]
    12	&	27.6	&	19.4	&	0.18	&	125	&	0.258	&	1.3	&	0.09	&	25	&	0.260	\\ [1ex]
    15	&	31.5	&	14.3	&	0.19	&	134	&	0.258	&	0.01	&	0.09	&	22	&	0.259	\\ [1ex]
    20	&	37.2	&	10.4	&	0.18	&	128	&	0.255	&	-	&	-	&	-	&	-	\\ [1ex]
    \rowcolor{mypink}
    \multicolumn{10}{c}{$\upsilon_{\rm ini}=0.4\,\upsilon_{\rm crit}$}	\\
    \noalign{\smallskip}	
    9	&	372	&	20.9	&	0.38	&	274	&	0.286	&	2.5	&	0.11	&	60	&	0.286	\\ [1ex]
    20	&	526	&	10.7	&	0.41	&	309	&	0.277	&	1.0	&	0.44	&	192	&	0.277	\\ [1ex]
    60	&	613	&	4.1	&	0.87	&	657	&	0.265	&	0.4	&	0.72	&	302	&	0.269	\\ [1ex]
    85	&	659	&	3.3	&	0.70	&	551	&	0.265	&	-	&	-	&	-	&	-	\\ [1ex]
    120	&	708	&	2.9	&	0.64	&	473	&	0.303	&	-	&	-	&	-	&	-	\\ [1ex]
    \rowcolor{mypink}
    \multicolumn{10}{c}{$\upsilon_{\rm ini}=0.7\,\upsilon_{\rm crit}$}	\\
    \noalign{\smallskip}	
    9	&	649	&	26.6	&	0.71	&	514	&	0.370	&	2.8	&	0.17	&	19	&	0.405	\\ [1ex]
    20	&	833	&	11.4	&	0.87	&	628	&	0.314	&	3.0	&	0.11	&	7	&	0.552	\\ [1ex]
    60	&	1080	&	4.3	&	0.82	&	565	&	0.313	&	0.3	&	0.10	&	12	&	0.557	\\ [1ex]
    85	&	1150	&	3.6	&	1.00	&	645	&	0.349	&	0.3	&	0.03	&	3	&	0.596	\\ [1ex]
    120	&	1230	&	3.0	&	0.86	&	512	&	0.366	&	-	&	-	&	-	&	-	\\ [1ex]
    \hline																	
    \hline																	
    \end{tabular}
    \label{table:models_details-2}
    \end{table}

As is well known \citep[see e.g.][]{Cassisi1993, Marigo2001, Sylvia2008}, the absence of any CNO elements in Pop~III stars leaves only $pp$-chains at the beginning to produce nuclear energy. The energy released by this process is however not enough to compensate for the losses at the surface and the star contracts (here we call this phase the {\it pp-contraction} phase). This occurs until temperatures are high enough to activate some synthesis of carbon by triple $\alpha$ reactions. From this stage on, the star reaches a radiative equilibrium in which the energy losses at the surface can be compensated for by the energy produced at the centre through nuclear reactions involving pp and CNO reactions ({\it pp-CNO-sustained} phase).

In the evolutionary tracks presented in Fig.~\ref{fig:HRDrot}, the transition between the {\it pp-contraction} phase and the {\it pp-CNO-sustained} phase is clearly visible in the case of the 9 and 20~{\msol} models. It corresponds to the points when the effective temperature begins to decrease as the luminosity increases (pink arrows in Fig.~\ref{fig:HRDrot}). In more massive stars, the {\it pp-contraction} phase is too short to be visible.

In the upper mass range, the models show inflexion points (clearly visible for the $\upsilon_{\rm ini}/\upsilon_{\rm crit}=0.7$ curves, less so for the slower-rotating models) during the MS phase. These inflexion points occur when the surface velocity reaches the critical limit. The mechanical mass loss slows down the surface below the critical value but its secular evolution will bring it back to the critical limit. This back-and-forth evolution produces the oscillations seen in effective temperatures along those tracks\footnote{During the short timescale of mechanical mass loss, the luminosity remains constant. Since when losing mass, the star loses angular momentum and becomes less oblate, for example, less deformed by rotation, its surface area, $S$, decreases and hence, through the relation $L=S\sigma T_{\rm eff}^4$, its effective temperature increases.}.

Our rapidly rotating models are slightly shifted to the red in the first part of the MS phase. Then, in the second part, they cross the slower-rotating track, reaching a higher luminosity. Except in the case of the 9~{\msol}, the rapidly rotating tracks show a turn-off at the end of the MS phase at lower effective temperatures than their slower-rotating counterparts. These two trends are well-known effects of rotation: the first one shows the hydrostatic effects of rotation \citep[see e.g.][]{Faul1968,Kip1970, MM1997}; the second one shows the mixing effects of rotation, mainly the enrichment in He of the radiative envelope \citep[][see also the discussion in Sect.~\ref{sec:32}]{Langer1992, MM2000} and the impact of the mixing on the convective core mass evolution \citep{Talon1997, Deupree1998, HL2000}.

From Fig.~\ref{fig:HRDrot}, we can see how changing the initial rotation impacts the nature of the star at a given evolutionary stage. As an example considering the 20~{\msol} model, the non-rotating model after the end of the core C-burning stage is a blue supergiant (BSG), while the rapidly rotating one is a red supergiant (RSG). We may thus expect very different SN light curves from the non-rotating and rapidly rotating models in case an explosion occurs at the time of the final core collapse \citep[see e.g.][]{Yoon2012, Aryan2023} (it should be noted that we do not expect any significant change in the surface conditions in the few tens of years that remain after the C-burning phase for the star to reach the final core-collapse phase).

Although the initial angular momentum of those models is high, the mixing is not efficient enough to make them evolve homogeneously \citep{Maeder1987}. Other rotating models \citep[see e.g.][]{Szecsi2015} show a chemically homogeneous evolution for an even slower initial rotation for masses above 26~{\msol} at $Z=0.0002$. Here of course we have metal-free Pop~III stars, which may explain in part this difference. Besides, the physics of rotation is different between the present models and those of, for example, \citet{Szecsi2015}. The latter include the Tayler-Spruit dynamo \citep{Spruit2002, Heger2005}. We would like to remind the reader of the fact that the term `rotating models' may actually correspond to very different physics and also to very different outputs. The results presented here correspond to one choice (non-magnetic models with specific expressions for the diffusion coefficients). The impact of other choices is explored in a dedicated work \citep{Nandal2023c}. It is still difficult to determine which kind of physics for the transport processes due to rotation are the most favoured. At solar metallicity, models with a more efficient angular momentum transport than those considered in this work are favoured. Indeed, transport accounting for the Tayler-Spruit dynamo produces a much better agreement with asteroseismic data of subgiants \citep{Moyano2023}\footnote{Note that these models include what has been called the Tayler-Spruit calibrated dynamo theory proposed by \citet{Eggen2022}. This implementation is different from the one accounted for in the models by \citet{Szecsi2015}.} than non-magnetic models like those presented here. However, we are exploring here a very special regime of initial composition (primordial stars) where it is not guaranteed that magnetic fields play a similar role. Also, the transport of the chemical species remains to be studied in more detail. Its efficiency (and thus its capability of driving a homogeneous evolution) depends mainly on the expression used for the horizontal shear turbulent coefficient. Thus, without more dedicated study, it is unfortunately still difficult to decide which types of models are the most representative of the bulk Pop~III star populations. We just retain here that the conditions for obtaining a homogeneous evolution heavily depend on the physics assumed for the transport of both the chemical elements and the angular momentum.

Some features of the Pop~III stellar models at the end of the core H- and He-burning phase are presented in Table~\ref{table:models_details-2}. Columns $1-2$ give the initial mass and the initial rotation on the ZAMS. Columns $3-6$ show, respectively, the duration of the MS lifetime, the values at the end of the core H-burning phase of the ratios between the surface to the critical velocity, the linear surface velocities at the equator, and the mass fractions of helium at the surface. The same quantities for the core He-burning phase are indicated in columns $7-10$.

From Table~\ref{table:models_details-2}, looking at column 10, we see that the non-rotating models show no increase in the surface He abundance between the end of the core H-burning phase and that of the core He-burning phase. This comes from the fact that the non-rotating models show no outer convective envelope during the core He-burning phase. They burn their helium in the core when the star is on the blue side of the HR diagram, and thus no important outer convective zone is present. Intermediate and outer convective envelopes appear (if any) only after the core He-burning phase. In contrast, rapidly rotating models show surface He-enrichment that is significantly increased with respect to the value reached at the end of the core H-burning phase. This is a consequence of the outer and intermediate convective zones that already appear during the core He-burning phase in those models.

Comparisons of the lifetimes of the core H- and He-burning phase for different initial rotations are given in columns 3 and 7 of Table~\ref{table:models_details-2}. The MS phase of the rapidly rotating 9~{\msol} model is about 50\% longer than the duration of the non-rotating one. Much smaller effects are obtained in higher masses. The increase is only 10\% for the 120~{\msol} model. Since the differences between the moderately rotating models \citep[see][]{LauraGrids} and the present rapidly rotating ones are modest during the MS phase for both the duration and the evolution in the HRD, we do not expect the results for ionisation fluxes based on the moderate rotation rate models \citep{LauraIoniz} to change much if rapidly rotating models are used instead.

\section{Primary nitrogen production in rotating models} \label{sec:5}

    \begin{figure*}
    \centering
    \includegraphics[width=1\textwidth]{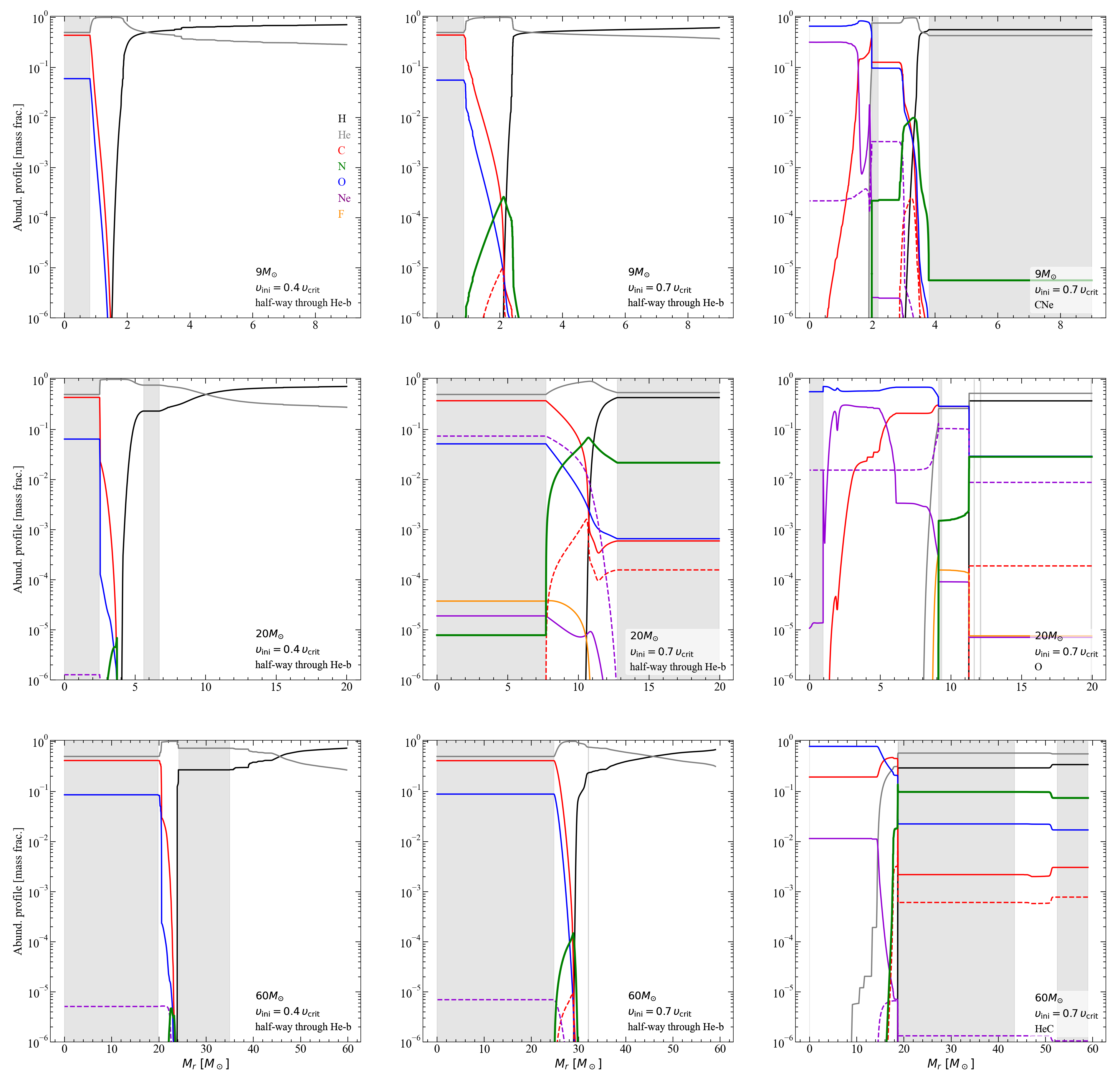}
    \caption{Chemical structure as a function of the Lagrangian mass coordinate in the interior of the Pop~III grids for $9 \msol \leqslant M_{\rm ini} \leqslant$ 60~{\msol}. The figures are at halfway through the core He-burning phase ($Y_c=0.5$) and at the last model that has been computed in different evolutionary stages. HeC is between the core He- and C-burning phase, CNe is the stage between the core C- and the Ne-burning phase, and O is during the core O-burning phase. The first column of the plots corresponds to the moderately rotating models with 40\% of the critical rotation, and the middle and last columns correspond to the rapidly rotating models with 70\% of the critical rotation. Each row refers to a different initial mass: the first row corresponds to 9~{\msol}, the second to 20~{\msol}, and the third to 60~{\msol}. Each coloured curve represents a different element, as is noted in the upper left plot. The light brown zones are the convective areas of each model. The moderately rotating models have been taken from \citet{LauraGrids}. The solid curves correspond to the most common isotope of an element, and the dashed curves to the second most common isotope.}
    \label{fig:chemYc500_endphase}
    \end{figure*}

    \begin{figure*}
    \centering
    \includegraphics[width=1\textwidth]{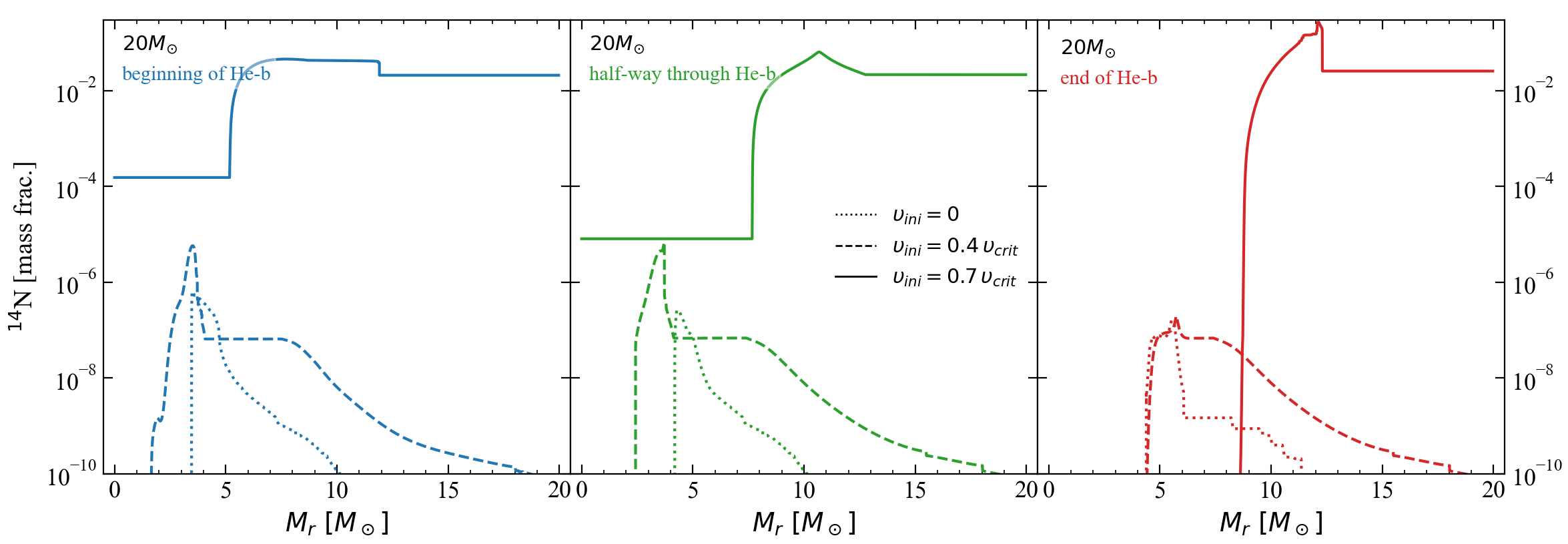}
    \caption{Abundance of $^{14}$N as a function of the Lagrangian mass coordinate during different stages of the core He-burning phase for the 20~{\msol} model with different initial rotations: zero (dotted curves), moderate (dashed curves), and fast (solid curves). The blue curves (left) correspond to the beginning of the core He-burning phase ($Y_c=0.9$), the green curves (middle) to the halfway point ($Y_c=0.5$), and the red curves (right) to the end of it ($Y_c=0$). The non-rotating and moderately rotating models were taken from \citet{LauraGrids}.}
    \label{fig:N14_Heb}
    \end{figure*}

    \begin{figure}
    \centering
    \includegraphics[width=0.47\textwidth]{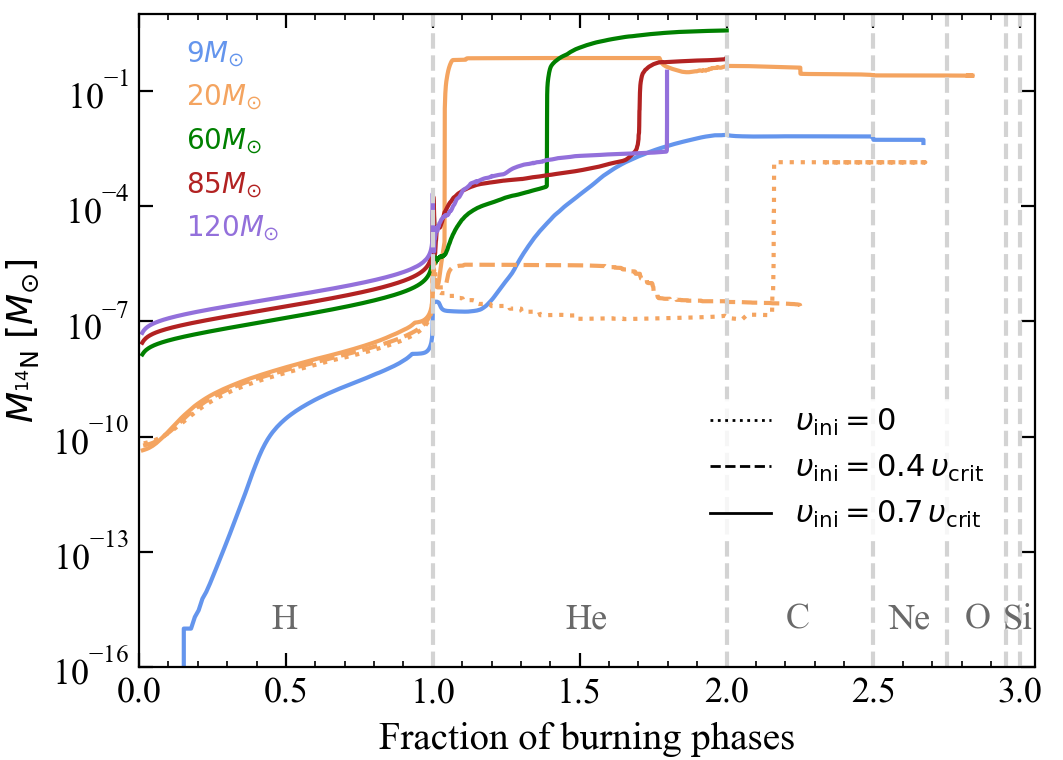}
    \caption{Evolution of the total mass of $^{14}$N inside the rapidly rotating stellar models during their evolution. For the 20~{\msol} model (orange), the non-rotating and moderately rotating models are also indicated by dotted and dashed curves. On the x-axis the fraction of burning phases is given, where the core H-burning phase is for 0$-$1, the core He-burning phase is for 1$-$2, and the advanced burning phases are for 2$-$3. The non-rotating and moderately rotating models were taken from \citet{LauraGrids}.}
    \label{fig:N_trel}
    \end{figure}

Figure~\ref{fig:chemYc500_endphase} shows the evolution of the chemical structure in different rotating models at different stages of their evolution. The light grey areas are the convective zones of the stars, while the white ones are the radiative zones.

The first column shows the structure in the middle of the core He-burning phase in models starting with a rotation equal to 40\% of the critical limit. We see that in those models some carbon and oxygen are brought by rotational mixing into the H-burning shell, inducing the production of a small amount of nitrogen (see e.g. the green curve bump at a mass of around 4~{\msol} in the 20~{\msol} model). Some of this nitrogen can diffuse into the He-burning core and be transformed into $^{22}$Ne (see the dashed purple curve into the He-burning core, e.g. below the mass coordinate 2.5~{\msol}). 
 
The panels in the middle column show models similar to those of the panels of the first column but now considering an initial rotation of 70\% of the critical one. Here we see that the bumps in nitrogen are much more prominent. More specifically, comparing the rapidly rotating models, we see that nitrogen dominates in the outer regions as we move towards to the surface in the 20~{\msol} model. This exceptional fraction of nitrogen can be attributed to the presence of the convective shell in this model, a feature absent in the 9 model and occurring in a later evolutionary phase in the 60~{\msol}. We can also see that $^{13}$C (see the dashed red curve) is produced in the H-burning shell (let us recall that $^{12}$C is transformed partially into $^{13}$C by the CNO cycle). Interestingly, in the 20~{\msol} model a significant quantity of $^{19}$F also appears in the He-burning core (this point will be discussed in a dedicated paper by Tsiatsiou et al. (in prep.)). Finally, $^{22}$Ne also reaches a high abundance in the He-core (see the dashed purple curve) as a result of the diffusion or mixing of nitrogen into the He-burning core.

The panels on the right show the same models as those of the middle panel but at a more advanced evolutionary stage, indicated in the caption of Fig.~\ref{fig:chemYc500_endphase}. We see that in general the abundance of nitrogen has increased in the outer layers of the star. This is because of the continuing build-up of this element in the H-burning shell due to the diffusion of carbon and oxygen into it. The appearance of convective zones can redistribute nitrogen over larger regions in the envelope. We note the very high level of nitrogen in the outer layers of the 20 and 60~{\msol} models.

Figure~\ref{fig:N14_Heb} allows us to see more clearly the impact of the initial rotation. These three panels compare the distribution of nitrogen as a function of the Lagrangian mass coordinate in the 20~{\msol} model with three different initial rotations at three different evolutionary stages of the core He-burning phase. We note that the nitrogen mostly builds up in the early phases of the core He-burning. Besides, in the rapidly rotating models ($\upsilon_{\rm ini}/\upsilon_{\rm crit}=0.7$), the increase is impressive. We see that even the central regions are enriched in nitrogen. This is a consequence of the very large intermediate convective zone that actually extends nearly to the centre (as can be seen in the lower panel of Fig.~\ref{fig:kippen-points20}), and thus brings nitrogen into very central regions. When He-burning restarts in the core, a new convective core appears that maintains, for a while at least, a high level of nitrogen abundance.

In Fig.~\ref{fig:N_trel}, the solid curves show how the integrated mass of nitrogen inside a model varies when the star evolves for different rapidly rotating models. The integrated mass at a given time, $t$, is given by $M_{^{14}\text{N}} = \int_0^{M_{tot}(t)} {}^{14}X(M_r,t) \, {\rm d}M_r$, where $M_{tot}(t)$ is the actual total mass of the star at the time, $t$, and ${}^{14}X(M_r,t)$ is the mass fraction of ${}^{14}$N at the Lagrangian coordinate, $M_r$, at time $t$.

We see that in the 9~{\msol} model, the increase in $M_{^{14}\text{N}}$ occurs progressively throughout the whole core He-burning phase. For the models with a higher initial mass, the increase is more step-like. As was just seen above, it occurs at the very beginning of the core He-burning phase in the 20~{\msol} model and at later and later core He-burning stages in more and more massive stars. In these models with a higher initial mass, convection plays an important role in connecting regions rich in both H- and He-burning products.

\begin{figure*}
    \centering
    \includegraphics[width=1\textwidth]{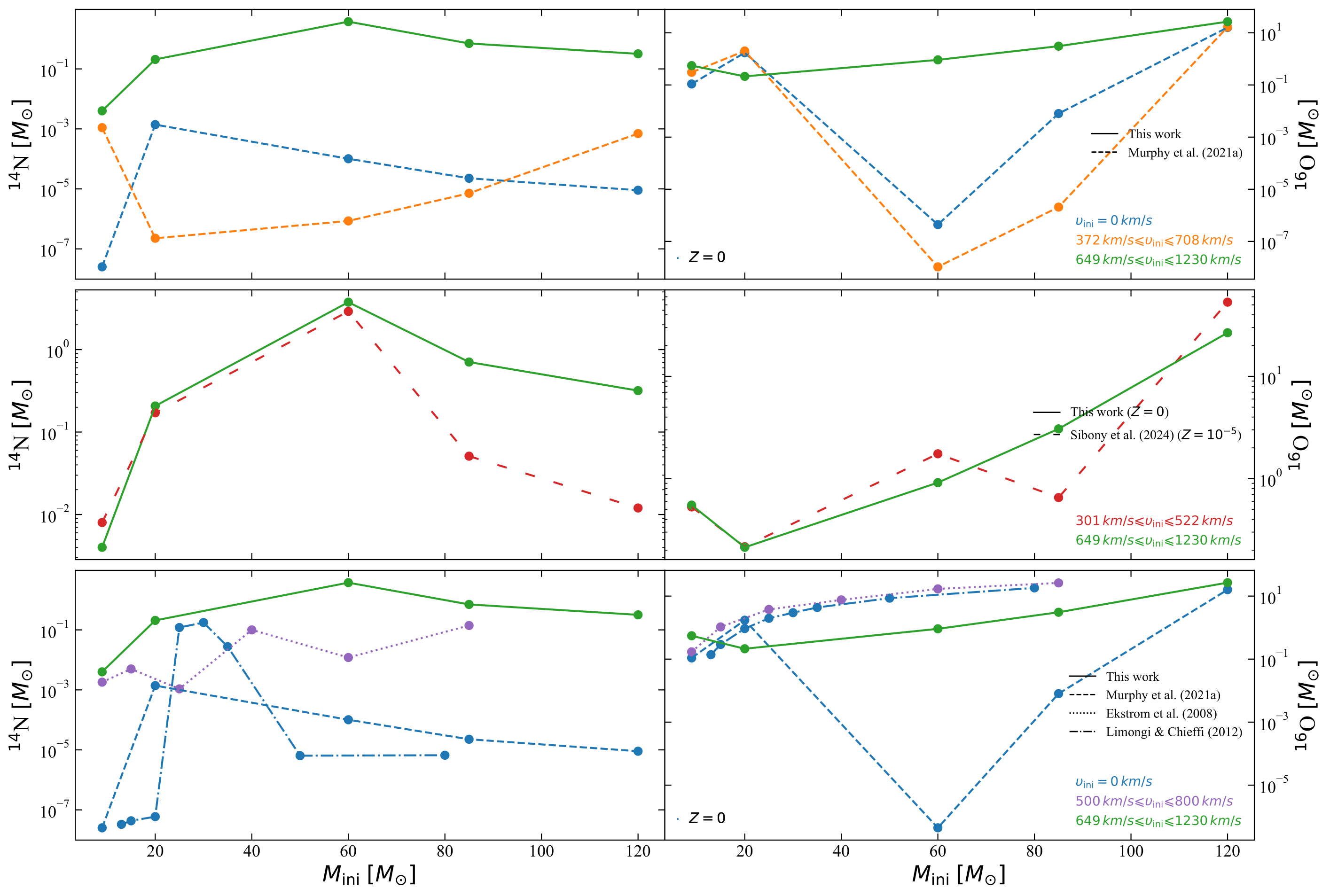}
    \caption{Stellar yields of nitrogen (left) and oxygen (right) for Pop~III and very low-metallicity ($Z=10^{-5}$) stars, for different initial rotations and from different authors \citep[][]{Sylvia2012,Limongi2012,LauraGrids,Sibonyzm5}. The ranges of initial equatorial velocities corresponding to the different coloured curves are indicated.}
    \label{fig:Yields_Patton}
    \end{figure*}

We have also plotted for the 20~{\msol} model the evolution of $M_{^{14}\text{N}}$ for the non-rotating and moderately rotating models (the dotted and dashed orange curves, respectively). Interestingly, even the non-rotating model shows a step-like increase in $M_{^{14}\text{N}}$ during the core C-burning phase. The moderately rotating model does not show such a feature but we cannot discard the fact that it may present one in a more advanced stage. We note that in non-rotating models such a connection between the He- and H-burning regions leading to significant primary nitrogen production seems to be restricted to a narrow mass domain \citep{Ekstrom2010, Limongi2012}, while rotational mixing not only boosts the primary nitrogen production in this mass domain but also extends the mass domain over which this process occurs across the whole mass range of stars considered here (see Fig. ~\ref{fig:Yields_Patton}).

\begin{figure}
    \centering
    \includegraphics[width=0.45\textwidth]{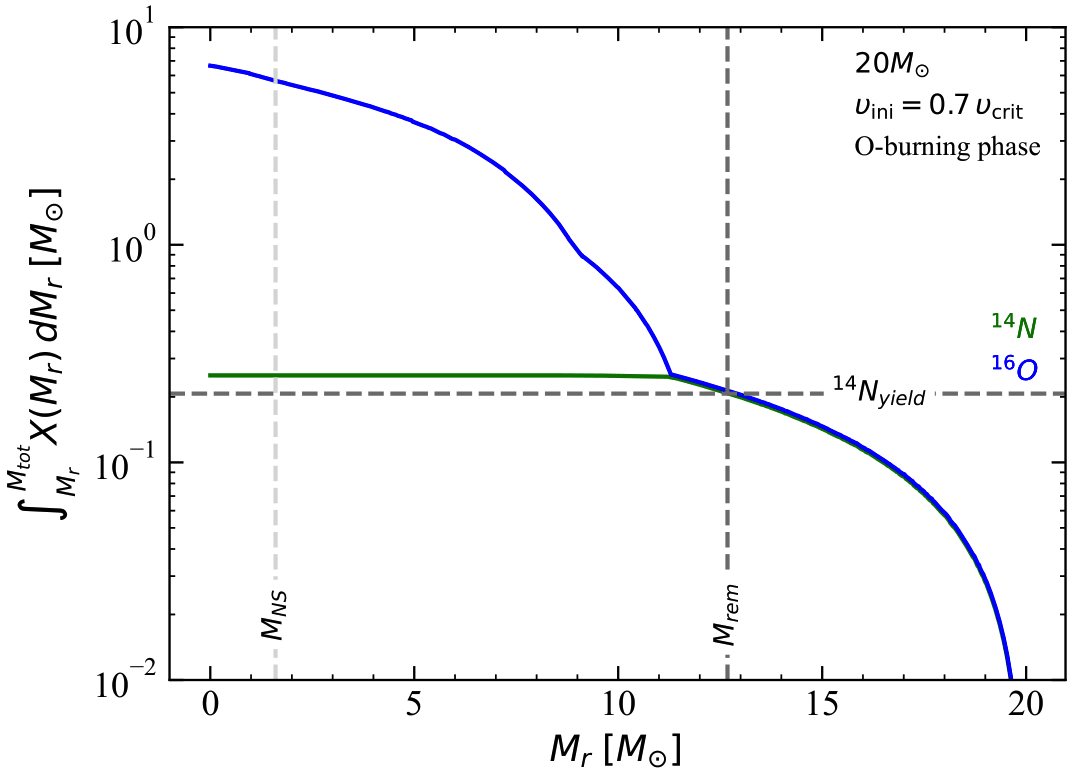}
    \caption{Variation in the mass of nitrogen and oxygen in the outer layers of the star as a function of the Lagrangian mass taken as the lower limit of the integration. This lower limit can be called the mass of the remnant. The case of the rapidly rotating 20~{\msol} model is shown. The vertical dashed grey line denotes the remnant mass adopted in the present paper, while the horizontal dashed grey line denotes the nitrogen yield. The dashed lighter grey line denotes the typical mass of an NS.}
    \label{fig:intNO}
    \end{figure}

Moreover, in Fig.~\ref{fig:intNO} we plot the integrated abundance of nitrogen (green) and oxygen (blue) as a function of the Lagrangian mass coordinate for the 20~{\msol} rapidly rotating model in its last evolutionary stage (during the core O-burning phase). We show in the plot where the remnant mass ($M_{\rm rem}$, dashed grey line) is located for the model, as well as the typical mass of a neutron star ($M_{\rm NS}$, dashed light grey line). It is clear that the selection of the remnant mass can have a strong impact on the oxygen yields. Typically, we see in Fig.~\ref{fig:intNO} that for the integrated mass of nitrogen, starting the integration in mass from the surface increases when considering still deeper layers until the Lagrangian mass around 11~{\msol} is reached. Below that mass, the nitrogen mass no longer increases because the nitrogen has been completely destroyed. Oxygen remains abundant in much deeper layers since O-burning only occurs in the very central parts of the star. As a consequence, the integrated mass of oxygen continues to increase when deeper layers are added to the mass ejected.

    \begin{table}
    \scriptsize
    \caption{Stellar yields of nitrogen and oxygen for different Pop~III and very metal-poor stellar models. The asterisk after the initial mass indicates that the model is still in the core He-burning phase. The Pop~III models with $\upsilon_{\rm ini}/\upsilon_{\rm crit}=$ 0, 0.2, and 0.4 are from \citet{LauraGrids} and the models with $Z=10^{-5}$ are from \citet{Sibonyzm5}.}
    \addtolength{\tabcolsep}{2pt}
    \begin{center}
    \begin{tabular}{cccccc}
    \hline											
    \hline \noalign{\smallskip}												
    $M_{\rm ini}$	&	$M_{\rm CO}$		&	$M_{\rm rem}$		&	Final	&	$^{14}$N	&	$^{16}$O		\\
    $[{\msol}]$	&	$[{\msol}]$	&	$[{\msol}]$	&	fate	&	$[{\msol}]$	&	$[{\msol}]$	\\
    \hline											
    \hline \noalign{\smallskip}											
    \rowcolor{operamauve}											
    \multicolumn{6}{c}{Z=0}											\\
    \rowcolor{mypink}											
    \multicolumn{6}{c}{$\upsilon_{\rm ini}=0$}											\\
    \noalign{\smallskip}											
    9	&	0.718	&	0.718	&	WD	&	2.52E-08	&	0.109	\\ [1ex]
    20	&	4.047	&	1.517	&	NS	&	1.39E-03	&	1.719	\\ [1ex]
    60	&	22.575	&	26.575	&	BH	&	1.01E-04	&	4.42E-07	\\ [1ex]
    85	&	31.768	&	35.768	&	BH	&	2.26E-05	&	0.008	\\ [1ex]
    120	&	52.887	&	34.640	&	PPISN	&	9.02E-06	&	15.969	\\ [1ex]
    \rowcolor{mypink}												
    \multicolumn{6}{c}{$\upsilon_{\rm ini}=0.2\,\upsilon_{\rm crit}$}	\\
    \noalign{\smallskip}											
    9	&	0.940	&	0.940	&	WD	&	2.89E-05	&	0.173	\\ [1ex]
    12	&	1.385	&	1.385	&	WD	&	1.27E-04	&	0.173	\\ [1ex]
    15	&	2.180	&	1.400	&	NS	&	2.88E-04	&	0.814	\\ [1ex]
    20*	&	19.99	&	19.99	&	BH	&	0.0	&	0.0	\\ [1ex]
    \rowcolor{mypink}												
    \multicolumn{6}{c}{$\upsilon_{\rm ini}=0.4\,\upsilon_{\rm crit}$}	\\
    \noalign{\smallskip}											
    9	&	0.895	&	0.895	&	WD	&	0.0011	&	0.304	\\ [1ex]
    20	&	3.981	&	1.785	&	NS	&	2.26E-07	&	2.006	\\ [1ex]
    60*	&	19.373	&	23.373	&	BH	&	8.54E-07	&	1.06E-08	\\ [1ex]
    85*	&	30.465	&	34.465	&	BH	&	7.15E-06	&	2.07E-06	\\ [1ex]
    120*	&	56.397	&	30.550	&	PPISN	&	0.0007	&	16.540	\\ [1ex]
    \rowcolor{mypink}												
    \multicolumn{6}{c}{$\upsilon_{\rm ini}=0.7\,\upsilon_{\rm crit}$}	\\
    \noalign{\smallskip}											
    9	&	1.389	&	1.389	&	WD	&	0.004	&	0.551	\\ [1ex]
    20	&	8.679	&	12.679	&	BH	&	0.207	&	0.213	\\ [1ex]
    60	&	14.442	&	18.442	&	BH	&	3.760	&	0.914	\\ [1ex]
    85	&	33.07	&	37.070	&	BH	&	0.706	&	3.068	\\ [1ex]
    120*	&	61.645	&	0.0	&	PISN	&	0.318	&	26.707	\\ [1ex]
    \rowcolor{operamauve}											
    \multicolumn{6}{c}{$Z=10^{-5}$}											\\
    \rowcolor{mypink}											
    \multicolumn{6}{c}{$\upsilon_{\rm ini}=0$}											\\
    \noalign{\smallskip}											
    9	&	1.009	&	1.009	&	WD	&	1.91E-05	&	0.123	\\ [1ex]
    20	&	3.748	&	7.748	&	BH	&	3.79E-05	&	-8.67E-05	\\ [1ex]
    60	&	20.925	&	24.925	&	BH	&	1.70E-04	&	-9.96E-05	\\ [1ex]
    85	&	31.622	&	35.622	&	BH	&	2.51E-04	&	0.001	\\ [1ex]
    120	&	47.977	&	41.620	&	PPISN	&	2.66E-04	&	6.292	\\ [1ex]
    \rowcolor{mypink}											
    \multicolumn{6}{c}{$\upsilon_{\rm ini}=0.4\,\upsilon_{\rm crit}$}	\\
    \noalign{\smallskip}											
    9	&	1.119	&	1.119	&	WD	&	0.008	&	0.529	\\ [1ex]
    20	&	2.980	&	6.980	&	BH	&	0.171	&	0.216	\\ [1ex]
    60	&	13.419	&	17.419	&	BH	&	2.902	&	1.750	\\ [1ex]
    85	&	37.358	&	41.358	&	BH	&	0.051	&	0.652	\\ [1ex]
    120	&	59.437	&	0.0	&	PISN	&	0.012	&	53.17	\\ [1ex]
    \hline											
    \hline											
	\label{table:yields_Patton}
    \end{tabular}
    \end{center}
    \end{table}

We use here the classical definition of the stellar yield of a given isotope. It is the quantity of the isotope that has been newly synthesised and ejected by the star. Since in the present models the mass lost in the course of the evolution is negligible, only the mass ejected at the time of the core collapse has to be accounted for. Thus, we only need to know the remnant mass to calculate the stellar yields. To obtain the final fate and remnant mass we used the results of \citet{Farmer2019, Patton2020}, following the method outlined in \citet{Sibony2023} and in Nandal, Sibony, \& Tsiatsiou (2024) for different CO core masses. The stellar yields were obtained by integrating the abundances of an element above the remnant mass and by subtracting the initial quantity of the element, as was indicated in \citet{Maeder1992_baryonic}. Of course for Pop~III stellar models, the mass of an ejected metal isotope is the stellar yield of that isotope.

The results are presented in Table~\ref{table:yields_Patton} for different metallicities and different initial rotations. The initial mass is given in column 1. In columns 2 and 3, the CO-core and the remnant masses are given, respectively. The CO-core mass corresponds to the mass inside the Lagrangian mass coordinate where the mass fraction of helium drops below $10^{-2}$ at the end of the core He-burning phase. We also provide the possible final fate of the stellar models in column 4. Lastly, the stellar yields of nitrogen and oxygen are given in columns 5 and 6, respectively.

The models were computed until the end of the core He-burning phase (some have been pushed further), so the yields that we obtain may still be affected by the more advanced stages of the evolution. However, the changes for the two isotopes considered here remain modest, and thus the yields obtained here are a good approximation of the yields that would be obtained from pre-SN models. Also, the bulk of nitrogen and oxygen is located sufficiently far away from the central conditions for them to be affected very little by explosive nucleosynthesis. 

Figure~\ref{fig:Yields_Patton} shows the stellar yields of nitrogen and oxygen. The upper panels show the yields for Pop~III models with different initial rotations. The non-rotating and moderately rotating models have been taken from \citet{LauraGrids}. We note that the stellar yields from \citet{LauraGrids} were re-calculated for this work, since we have a different method of calculating the remnant mass. In \citet{LauraGrids} the remnant mass was calculated according to \citet{Maeder1992_baryonic}. The middle panels show the yields from the rapidly rotating Pop~III models, together with the moderately rotating metal-poor models ($Z=10^{-5}$) from \citet{Sibonyzm5}. The two lower panels show the yields for Pop~III models with different initial rotations and from different authors. The non-rotating models have been taken from \citet{Limongi2012} and \citet{LauraGrids}, and the moderately rotating from \citet{Sylvia2008}. We discuss these two lower panels in Sect.~\ref{sec:62}.

The production of $^{14}$N is boosted in the rapidly rotating models, as can be seen by comparing the solid green curve with the dashed blue (no rotation) and orange (moderate rotation) curves. The moderately rotating models show smaller yields of nitrogen than the non-rotating ones in the mass range between about 10 and 87~{\msol}. As is explained in Sect.~\ref{app:A3}, this results from very different behavior of the H-burning shell in the two models. Actually, in the non-rotating models, the structure shows that the H- and the He-burning shell are much nearer to each other than in the rotating model. This favours mixing between these shells in the non-rotating model after the core He-burning phase.

Regarding the comparison between the Pop~III models and those with $Z=10^{-5}$, we see that the rapidly rotating Pop~III stars may contribute even more than the moderately rotating $Z=10^{-5}$ models (see middle left panel). The $Z=10^{-5}$ models in the high mass range lose significant amounts of mass by stellar winds. As a numerical example, the 85~{\msol} model at $Z=10^{-5}$ loses about 30~{\msol}. This has a strong impact on the structure, notably making the H-shell less active and more distant from the He-burning region, and thus making primary nitrogen production less efficient. In Fig.~\ref{fig:intNO}, we show how the yield of nitrogen would vary when considering different remnant masses for the 20~{\msol} model. For instance, for the remnant mass given in Table~\ref{table:yields_Patton}, the yield is given by the co-ordinates of the intersection between the nitrogen curve and the vertical dashed line labelled $M_{\rm rem}$ in Fig.~\ref{fig:intNO}. Considering lower values for the remnant mass would not change the nitrogen yield much. Considering larger values would, on the contrary, significantly impact the yield, even suppressing it completely if the whole final mass of the star were to remain locked into the stellar remnant black hole.

The yields for the main isotope of oxygen show significant differences between the rapidly rotating non-rotating, and moderately rotating models. Most of these differences reflect a complex interaction between the effects of rotation on the size of the CO-core and the process of computing the remnant mass. Changing the remnant mass may strongly impact the yields of oxygen. Looking at Fig.~\ref{fig:intNO}, we see that in the case in which the rapidly rotating 20~{\msol} model produces a 12.7~{\msol} black hole or a 1.6~{\msol} NS, the yield of oxygen varies from a value of 0.2~{\msol} up to a value of nearly 6~{\msol}. This high sensitivity of the oxygen yields on the remnant mass explains the very large difference between the yields of the non-rotating or moderately rotating 60~{\msol} models and the rapidly rotating one. In the rapidly rotating model, a smaller CO-core mass is produced (14.4~{\msol} instead of 22.6~{\msol} in the non-rotating model). This is a result, as was already explained above, of the extended intermediate convective zone attached to the H-burning shell when a CNO shell boost occurs. This has the consequence of reducing the He- and then the CO-core mass and finally the remnant mass in the rapidly rotating model. It has a dramatic impact on the oxygen yields, passing from a value equal to $4.4\times 10^{-7}$~{\msol} in the non-rotating model to a value equal to 0.91~{\msol} in the rapidly rotating one. This illustrates the great sensitivity of the oxygen yield to the final fate of a star, a point that was already discussed in detail by \citet{Maeder1992_baryonic}. Given this, and since the final fate of massive stars is still a topic associated with great uncertainties, we provide on our \href{https://www.unige.ch/sciences/astro/evolution/en/database/}{webpage}\footnote{\url{https://www.unige.ch/sciences/astro/evolution/en/database/}} the structure of our last computed models, so that other remnant masses can be considered to obtain the yields.

In the middle right panel of Fig.~\ref{fig:Yields_Patton}, the oxygen yields for the $Z=10^{-5}$ models with $\upsilon_{\rm ini}/\upsilon_{\rm crit}=0.4$ are compared with the Pop~III oxygen yield from models with $\upsilon_{\rm ini}/\upsilon_{\rm crit}=0.7$. Some masses show very similar values, and others significant differences, without any systematic effects. Again, here the behaviour is very dependent on the way the remnant masses were computed.

\section{Discussion} \label{sec:6}

In this section, we discuss three points: how the primary nitrogen production depends on the metallicity, considering the whole range from $Z=0$ to $Z=0.020$; how the present yields compare with other works; and how our yields impact the variation of N/O as a function of O/H when incorporated into a simple chemical evolution model. 

\subsection{The metallicity dependence} \label{sec:61}

    \begin{figure}
    \centering
    \includegraphics[width=0.45\textwidth]{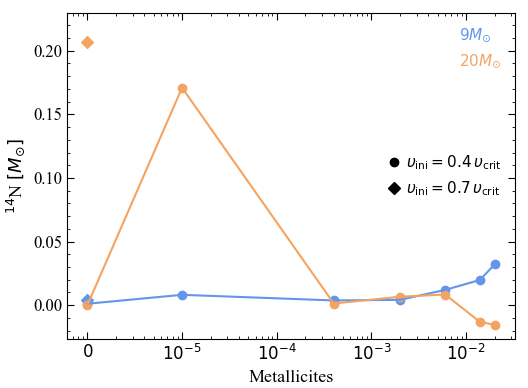}
    \caption{Stellar yields of nitrogen for the 9 (blue) and 20~{\msol} (orange) models at different metallicities ($0 \leqslant Z \leqslant 0.020$). The circles connected by the solid curve correspond to models with an initial rotation of $\upsilon_{\rm ini}=0.4\,\upsilon_{\rm crit}$ and the diamonds ones with $\upsilon_{\rm ini}=0.7\,\upsilon_{\rm crit}$. The moderately rotating models are from \citet[][$Z=0$]{LauraGrids}, Sibony et al. (2014, $Z=10^{-5}$), \citet[][$Z=0.0004$]{Groh2019}, \citet[][$Z=0.002$]{Cyril2013}, \citet[][$Z=0.014$]{Sylvia2012}, \citet[][$Z=0.006$]{Eggenberger2021}, and \citet[][$Z=0.020$]{Yusof2022}.}
    \label{fig:all_metal}
    \end{figure}

    \begin{figure}
    \centering
    \includegraphics[width=0.49\textwidth]{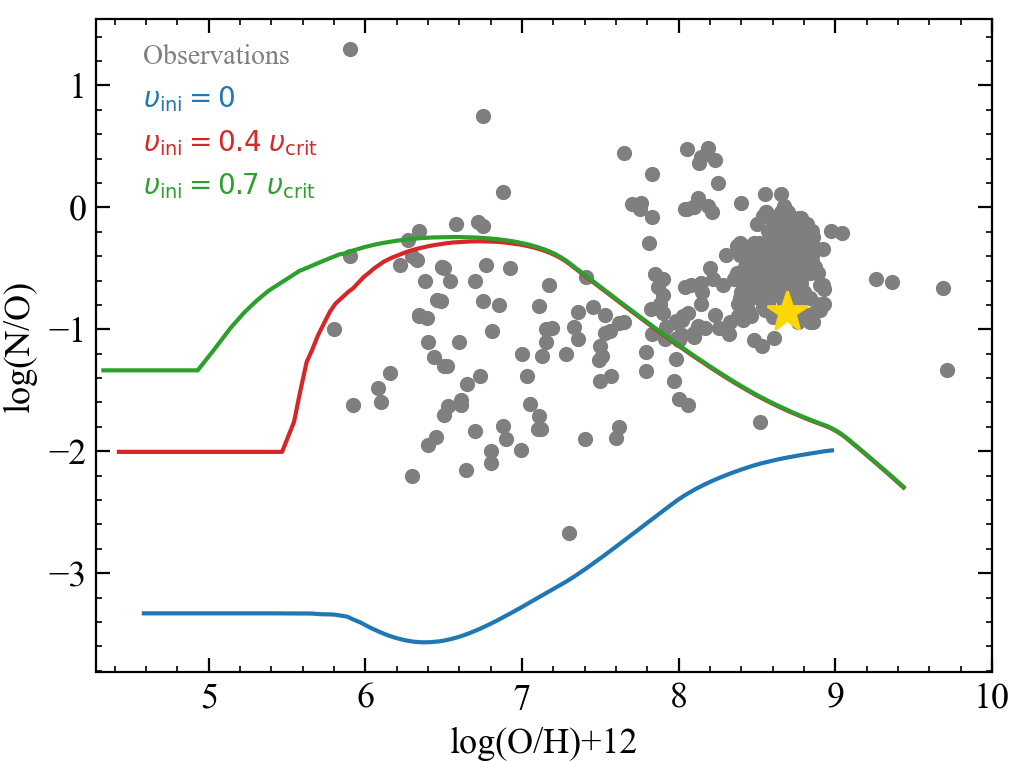}
    \caption{Evolution of N/O as a function of O/H ratios for stellar models in the metallicity range of $0 \leqslant Z \leqslant 0.020$ (coloured curves) and observed abundances of the Milky Way's stars (grey points). Yields from models with different initial rotations were used (see Sect.~\ref{sec:62}). The observed data were taken from the SAGA database \citep{Suda2008, Suda2011, Yamada2013}. The yellow star represent the Sun's abundances from \citet{Asplund2009}.}
    \label{fig:Yutaka}
    \end{figure}

Figure~\ref{fig:all_metal} shows how the stellar yields vary as a function of the initial metallicity for the 9 (orange) and 20~{\msol} (blue) models. The solid curves show the results when the initial rotation is $\upsilon_{\rm ini}/\upsilon_{\rm crit}=0.4$ at every metallicity, while the diamonds indicate the results when fast rotation is considered for $Z=0$.

Let us first discuss the case of the 20~{\msol} models. Schematically, we see that for the metallicity range between about 0.0004 and 0.020 with $\upsilon_{\rm ini}/\upsilon_{\rm crit}=0.4$, the yields in nitrogen remain modest. They are even negative in the upper metallicity range. For those metallicities, nitrogen is mainly produced by the transformation of the initial carbon and oxygen abundances and nearly no nitrogen is produced by the diffusion of carbon and oxygen produced in the He-core into the H-burning shell. Yields are negative at high $Z$ because the mass of the ejecta where nitrogen is destroyed (typically, matter processed by He-burning) is larger than the mass of the ejecta where nitrogen is produced.

It is only in the metallicity range between 0 and 0.0004 that primary nitrogen is produced thanks to rotational mixing. Starting from $Z=0$, the increase in the yield with $Z$ is mainly due to rotational mixing, which is favoured in slightly metal-rich models. As was explained before, the differential rotation that drives the mixing increases when the metallicity increases. Beyond the peak, the decrease in the yield when the metallicity increases further is due to the larger and larger distance between the He- and H-burning zones. The 20~{\msol} rapidly rotating model is available only for $Z=0$. For that model, the primary nitrogen production reaches a level that is larger than the peak value shown at a metallicity of $Z=10^{-5}$ with a moderate initial rotation.

The 9~{\msol} Pop~III models with $\upsilon_{\rm ini}/\upsilon_{\rm crit}=0.4$ do not show any strong peak between $Z=0$ and 0.0004. Primary nitrogen production in this metallicity range depends sensitively on the initial mass. It is interesting to note here that the central temperatures during the H-burning phase are smaller in less massive stars. Thus, decreasing the mass, everything else being kept constant, also tends to increase the change in entropy between the He- and H-burning zones, making mixing more difficult between the two zones. 

\subsection{Comparisons with other works} \label{sec:63}

Comparing the nitrogen yields in Fig.~\ref{fig:Yields_Patton} for the non-rotating models computed with {\genec} by \citet{LauraGrids} (dashed blue curve) with those of \citet{Limongi2012} (dash-dotted blue curve), one can note the following points. First, our yields for the 9~{\msol} model are in line with those of \citet{Limongi2012}. Qualitatively, we also obtain a bump that begins around 20~{\msol}. Since our grid does not consider the same initial masses, it is difficult to say more at the moment. We can just add that our yield for the 85~{\msol} is also in line with the yield given by \citet{Limongi2012}. 

If we compare the nitrogen yields by \citet{Ekstrom2010} for rotating models computed with {\genec} with those presented in this work, we obtain significant differences. This is in part due to the faster initial rotation considered here but also due to the fact that in the present models we used different diffusion coefficients for rotational mixing.

The differences between the yields in oxygen are mainly due to the different methods used to estimate the remnant mass. It is interesting to note that yields of \citet{Sylvia2008} for rotating models and those of \citet{Limongi2012} for non-rotating Pop~III stars are very similar. Actually, in these two papers, the authors used remnant masses that are in general much smaller than those in this work. This gives larger oxygen yields that do not differ much between the rotating and non-rotating models.

These yields differ significantly from those obtained with the rapidly rotating models computed here and even more from the non-rotating models of \citet{LauraGrids}. As was explained already, this comes from the use of the method of \citet{Patton2020} to estimate the remnant mass.

\subsection{Impact on the early chemical evolution of galaxies} \label{sec:62}

Figure~\ref{fig:Yutaka} shows the impact of the yields with different initial rotations in a simple galactic chemical evolution (GCE) model. Here, we computed the GCE with the one-zone closed-box model compiled in the chemical evolution library \citep[\textsc{celib},][]{Saitoh2017, Hirai2021}. This model adopts the initial mass function of \citet{Kroupa2001} for 0.1$-$120~{\msol} and stellar lifetimes in \citet{Portinari1998}. We assume stars are constantly formed for 13~Gyr from a gas cloud of $10^{11}$~{\msol} with a star formation efficiency of 0.08~Gyr$^{-1}$. We note that rapidly rotating models ($\upsilon_{\rm ini}/\upsilon_{\rm crit}=0.7$) were computed only for Pop~III models. Therefore, we adopt the rapidly rotating model for $Z$ = 0 and the rotating model with $\upsilon_{\rm ini}/\upsilon_{\rm crit}=0.4$ from \citet{Sylvia2012, Cyril2013, Groh2019, Eggenberger2021, Yusof2022}; Sibony et al. (2024) for metallicities above zero in the green curve. The red curve corresponds to the models above and to the models from \citet{LauraGrids} with $\upsilon_{\rm ini}/\upsilon_{\rm crit}=0.4$ for the metallicity $Z=0$. Lastly, the blue curve corresponds to non-rotating models from all the works above. Observed data of the Milky Way's stars were taken from the SAGA database \citep{Suda2008, Suda2011, Yamada2013}. We excluded CEMP stars with [C/Fe] $>+$0.7, which might be affected by binary mass transfer.

We see that, depending on the initial rotation, very different values of the initial plateau for $\log({\rm O/H})+12<5$ are obtained, covering more than two orders of magnitude. Because of the increased nitrogen production in the model with $\upsilon_{\rm ini}/\upsilon_{\rm crit}=0.7$, the $\log({\rm N/O})$ value at $\log({\rm O/H})+12 = 5$ is predicted to be $-$1.3, while the non-rotating model predicts $\log({\rm N/O}) = -3.3$. The peaks of the $\log({\rm N/O})$ seen around $\log({\rm O/H})+12 = 6.3$ in the rotating models are caused by the maximum yields of $^{14}$N at $Z=10^{-5}$ shown in Fig. \ref{fig:all_metal}. Most stars with $\log({\rm O/H})+12 < 7$ are within the N/O ratios predicted in models with $\upsilon_{\rm ini}/\upsilon_{\rm crit}=0,\,0.4$ and 0.7. The increasing trend seen in the observations for $\log({\rm O/H})+12 > 7$ could be explained by the contribution of asymptotic giant branch (AGB) stars, which are not included in this model \citep[e.g.][]{Chiappini2006}. This example is just to illustrate that indeed the increase in primary nitrogen production obtained in the present models can have a significant impact at very low metallicities. In a complementary paper, we shall publish the yields obtained from our stellar grids of models covering the whole metallicity domain and further works including these new yields will allow us to predict the evolution of other elemental ratios.

\section{Conclusion} \label{sec:7}

We have explored the impact of fast rotation on the evolution of Pop~III stars with some discussion of their chemical feedback. The main conclusions are briefly synthesised below:

\begin{enumerate}
\item We obtain that for initial masses between 9 and 60~{\msol}, primary nitrogen production in Pop~III stellar models beginning their evolution with 70\% of the critical velocity on the ZAMS is similar to the primary nitrogen production in models beginning their evolution with 40\% of the critical velocity on the ZAMS at a metallicity of $Z=10^{-5}$.
\item The appearance of the extended intermediate convective zone induced by the injection of carbon and oxygen produced in the He-burning region into the H-burning shell by diffusion or convection may have a dramatic impact on the structure and the evolution of the star. This intermediate convective zone may quench the He-burning in some models in the core for a while or at least significantly reduce the He-burning core \citep[a process called CNO shell boost, as in][]{Sylvia2008}. This has an impact on the nature of the stellar remnant and on nucleosynthesis.
\item Models with a mass equal to or larger than 60~{\msol} spend more than half of their MS lifetime at the critical velocity limit. However, the mass lost mechanically remains very modest; the most massive model (120~{\msol}) loses 2\% of its initial mass at the end of its evolution.
\item With the physics used here, models starting with 70\% of the critical velocity on the ZAMS show no evidence of following a chemically homogeneous evolution trend. Actually, their tracks during the MS phase follow the classical redward evolution with the models rotating faster, showing a greater extension towards lower effective temperatures than slower rotators. The impact of fast rotation on the ionisation outputs appears to be modest in the present models.
\item Depending on the initial rotation at the ZAMS, a Pop~III star can end its stellar lifetime as a BSG or an RSG. The rapidly rotating models will be RSGs, as opposed to the BSGs from the slower-rotating models (for $M_{ini}>20$~{\msol}).
\end{enumerate}

This work illustrates the fact that the physics of rotational mixing is a key point that needs to be resolved in order to predict stellar yields, especially for the first generations of stars. The physics of rotation is in our view as important as the impact of close binary evolution. Indeed, if mass transfer in close binaries can strongly impact the evolution of stars, primary nitrogen production is essentially a question related to the internal mixing. Rotation appears to be an excellent candidate for what is driving this mixing (we note that close binary evolution and rotation are tightly intertwined through the exchange of angular momentum between orbits and spins of stars). An interesting feature of the rotational mixing is that it shows a dependence on the metallicity, which is quite in line with the fact that important sources of primary nitrogen are needed only at low metallicities \citep[see e.g. the discussion in][]{Chiap2005}.

This work also underlines the importance of the initial rotation of Pop~III stars. Moreover, even if we had a good knowledge of the rotational velocity distribution in the first Pop~III stars, significant uncertainties persist regarding the physics of rotation. Similar work to that discussed here should be done with other approaches to describe the transport of angular momentum and of chemical species by rotation and also by other types of instabilities (improvement in the physics of convection, internal waves, and magnetic instabilities). There is some hope that testing stellar physics on stars that we can observe in the present-day Universe will allow us to constrain these mechanisms in a still-tighter way. Then one will be able to assume that this physics also applies to Pop~III stellar models. At the moment, and likely for a very long time to come, there is no way of directly measuring the rotation periods of Pop~III stars from observations. Thus, the only way to get information about that is to observe the consequences of stellar models computed with various initial rotations, ideally with different physics for the transport mechanisms, on the early chemical evolution of galaxies (as the origin of the chemical composition of CEMP stars and of the bulk of the galactic halo stars). A possible exception could be provided by cases of extreme flux amplification due to strong gravitational lensing \citep[e.g.][]{Zackrisson2015}. Recently, such extreme lensing has been invoked to explain the Earendel stellar source \citep{Welch2022}, which has a non-negligible probability of being Pop~III, although it is more likely to be a metal-poor, massive Pop~II star \citep{Schauer2022}. Another channel for getting information on the evolution of the first massive star generations is through predictions of the rate of various types of SNe \citep[see the reviews by][and references therein]{Frebel2015, MMb2017}. Recently, nitrogen-rich high-redshift galaxies have been observed \citep{Oesch2016}. The implications of the present rapidly rotating models in this context are discussed in another paper (Nandal et al., in prep.).


\begin{acknowledgements}
ST, YS, DN, SE and GM have received funding from the European Research Council (ERC) under the European Union's Horizon 2020 research and innovation programme (grant agreement No 833925, project STAREX). LS and SE have received support from the SNF project No 212143. EF has received support from the SNF project number 200020\_212124. YH has received JSPS KAKENHI Grant Numbers JP22KJ0157, JP20K14532, JP21H04499, JP21K03614, JP22H01259. BL is supported by the Royal Society University Research Fellowship.
\end{acknowledgements}


\bibliographystyle{aa_url}
\bibliography{arXiv}


\begin{appendix}
\section{Additional figures} \label{app:A}

\subsection{Case of the rapidly rotating 120~{\msol} model} \label{app:A1}

\begin{figure}[h]
    \centering
    \includegraphics[width=0.49\textwidth]{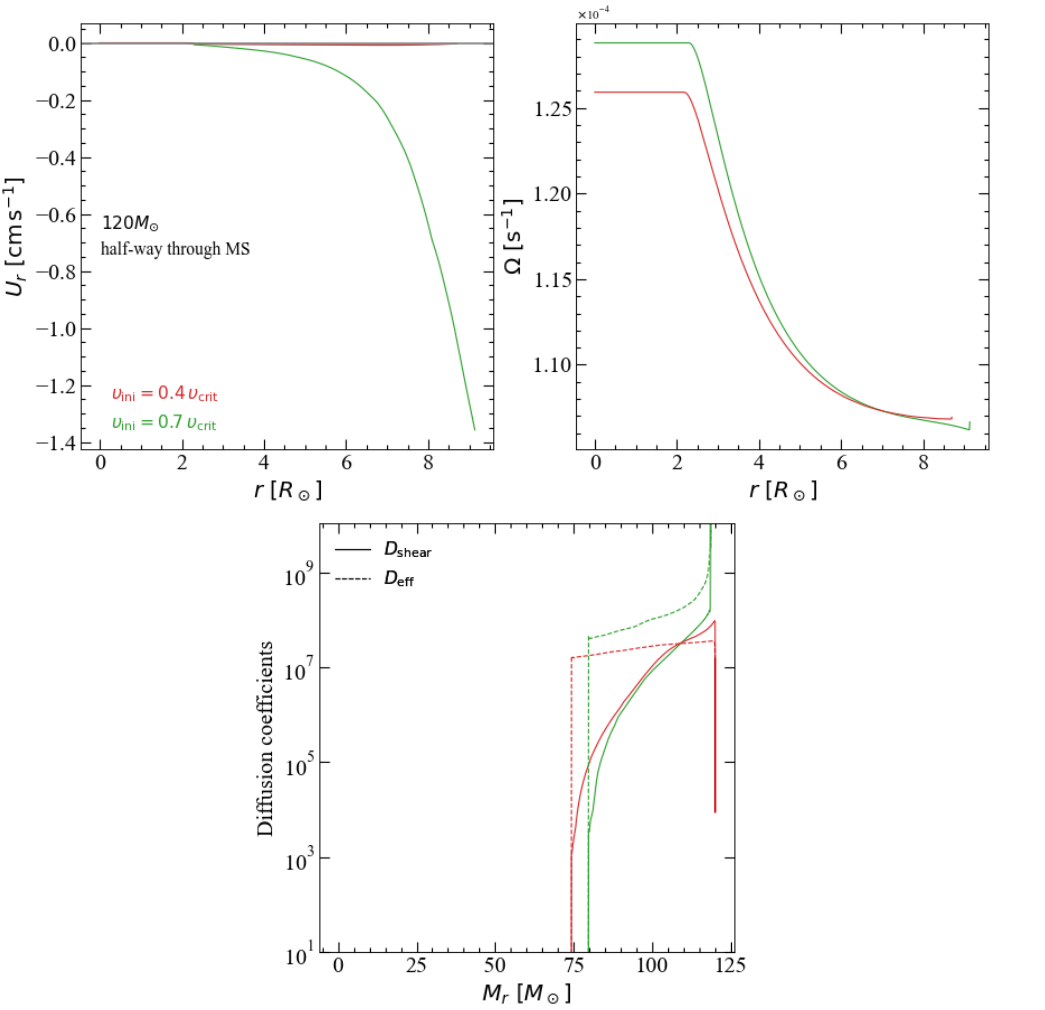}
    \caption{Variation of $U_{\rm r}$ ({\it upper left panel}), $\Omega$ ({\it upper right panel}), $D_{\rm shear}$ (solid curves) and $D_{\rm eff}$ (dashed curves) ({\it lower panel}) as a function of the Lagrangian mass coordinate in the Pop~III 120~{\msol} models with different initial rotations. The moderately rotating model was taken from \citet{LauraGrids}.}
    \label{fig:120-Xc035}
    \end{figure}

Figure~\ref{fig:120-Xc035} compares the variation of $U_{\rm r}$, $\Omega$, $D_{\rm shear}$ and $D_{\rm eff}$ as a function of the Lagrangian mass during the core H-burning phase when the mass fraction of hydrogen $X_{\rm c}$ is equal to 0.35 for the 120~{\msol} models with a moderate (red curves) and a fast (green curves) initial rotation. The rapidly rotating model shows stronger meridional currents (see the green curve in the upper left panel), which transport angular momentum from inside to outside. The value of $D_{\rm eff}$ is larger by many orders of magnitudes than $D_{\rm shear}$ in the radiative envelope and governs the mixing of the elements in that region.

\subsection{CNO shell boost in a Pop~III 60~{\msol} model} \label{app:A2}

    \begin{figure}
    \centering
    \includegraphics[width=0.45\textwidth]{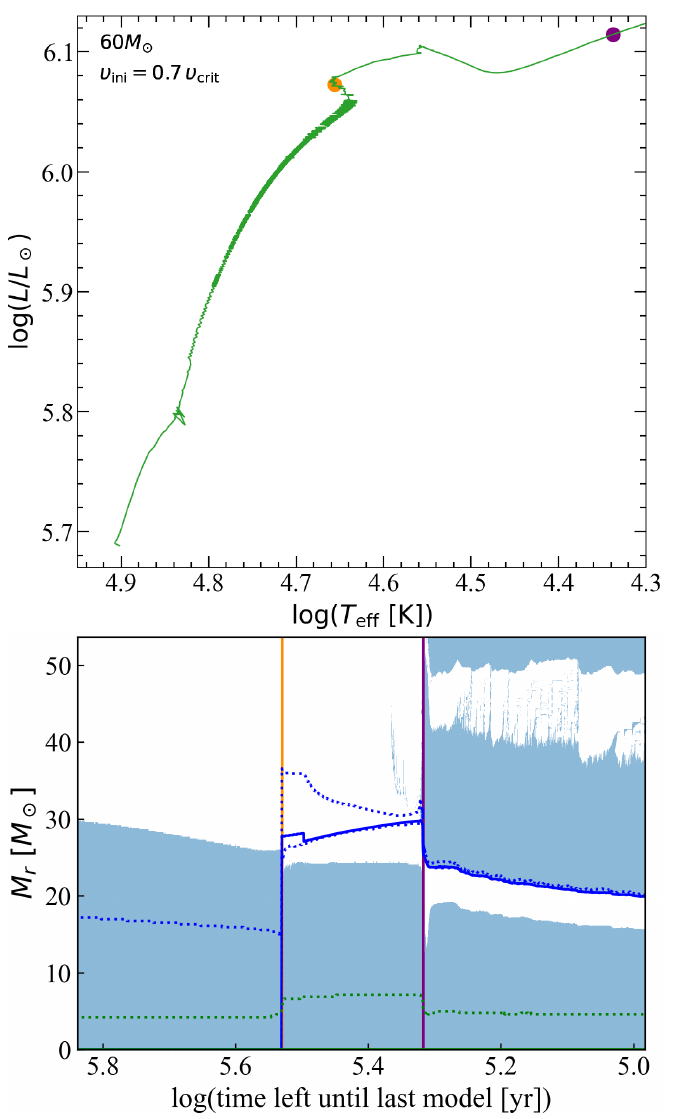}
    \caption{Comparing stages for the 60~{\msol} rapidly rotating model as in Fig.~\ref{fig:kippen-points20}.}
    \label{fig:kippen-points60}
    \end{figure}
    
The upper panel of Fig.~\ref{fig:kippen-points60} shows the HRD of the rapidly rotating Pop~III 60~{\msol} model. Coloured points are overplotted. The orange point indicates when the core He-burning phase begins, while the purple point indicated the stage when an extended intermediate convective zone appears as a result of injections of carbon and oxygen from the He-burning core to the H-burning shell. Meanwhile, the convective He-burning core is reduced. Those points correspond to the vertical coloured lines in the lower Kippenhahn plot, respectively. 

\subsection{Reason why rotating models may produce less nitrogen than non-rotating ones.} \label{app:A3}

    \begin{figure*}
    \centering
    \includegraphics[width=0.45\textwidth]{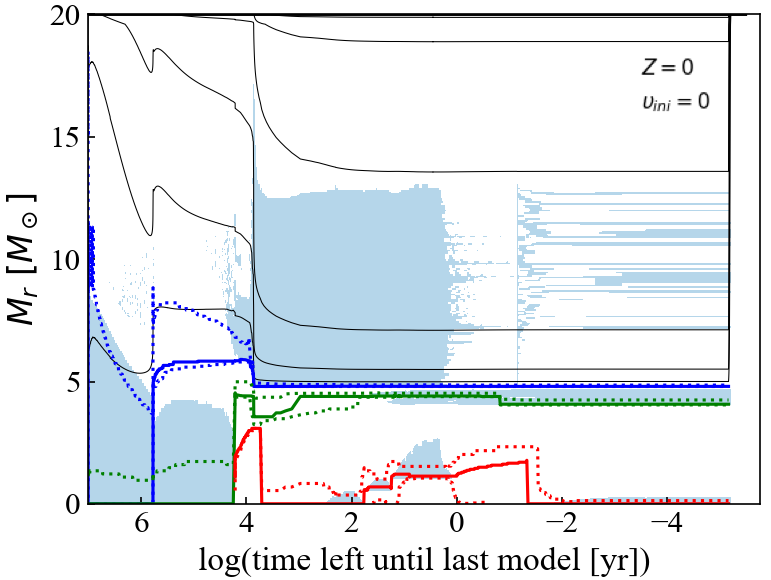}
    \includegraphics[width=0.45\textwidth]{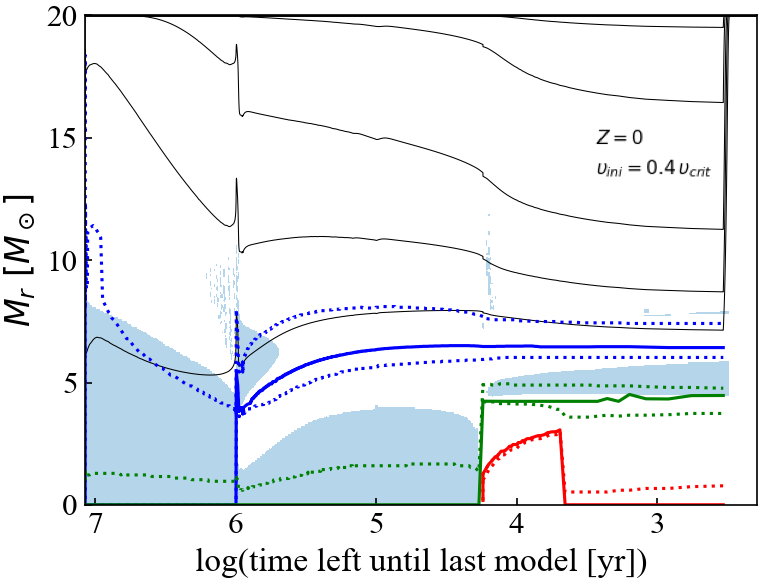}
    \caption{Kippenhahn diagrams for the 20~{\msol} Pop~III models with zero (left panel) and moderate (right panel) rotation.}
    \label{fig:kippen-20M}
    \end{figure*}

In Fig.~\ref{fig:kippen-20M}, we have an example of 20~{\msol} Pop~III models with zero and moderate rotation. We notice that the H- and He-burning shells are very close to each other in the non-rotating model. More precisely, in the non-rotating model, the mass separating the H- to the He-burning shell is a fraction of 1~{\msol} during the core C-burning phase. While, in the moderately rotating model, during the same phase, the mass separating the H- and He-burning shell is more than 2~{\msol}. This is in our view an indirect consequence of the fact that in the moderately rotating model (right panel), a small convective zone occurs above the H-shell at the very beginning of the core He-burning phase (due to the rotational mixing some C and O migrate into the H-shell and boost its energy production). This has for an effect to make the mass of the convective He-core smaller during the first part of the He-burning phase. If the H-shell is replenished in hydrogen by the convective shell, it stays longer in deeper regions of the star, so it migrates slowly towards outer regions. In the non-rotating model, the H-burning shell at the beginning of the core He-burning phase migrates much faster outwards than in the rotating model (the shell is not replenished by the intermediate convective zone and thus progresses more rapidly outwards). This makes the He-core to also increases more rapidly in mass. As a result of this behaviour the two shells, just after the end of the core He-burning phase are separated by less mass in the non-rotating model than in the rotating one. Compare, for instance, in the two Kippenhahn diagram (see Fig.~\ref{fig:kippen-20M}) the mass between the blue and green solid lines at a time 4. Due to the greater proximity in the non-rotating model, a CNO shell boost occurs. As a result, we see that the H-shell shifts to a position at a smaller Lagrangian mass coordinate. So we see how complex can be this behaviour. In the rotating model a modest early CNO boost occurs that create conditions that disfavour any further CNO shell boost. In the rotating model the early CNO shell boost does not occur in absence of any rotational mixing, but it creates conditions more favourable for a CNO boost occurring at the end of the core He-burning phase.

Moreover, at the end of core He-burning phase, the moderately rotating mode has higher nitrogen yield (approximately the moderately rotating has $^{14}{\rm N}=3.2\times10^{-7}$, while the non-rotating has $^{14}{\rm N}=1.2\times10^{-7}$). Therefore, we see that those yields on the Fig.~\ref{fig:Yields_Patton} are a result after the core He-burning phase.

\end{appendix}

\end{document}